\documentclass[reprint,amsmath,amssymb,aps,showpacs,floatfix,byrevtex,longbibliography]{revtex4-1}
\usepackage{amsmath}
\usepackage{amssymb}
\usepackage{amstext}
\usepackage{amsopn}
\usepackage{amsfonts}
\usepackage{amsxtra}
\usepackage[english]{babel}
\usepackage{graphicx}
\usepackage{float}
\usepackage{bm}
\usepackage{multirow}
\usepackage{dcolumn}
\usepackage{color}
\usepackage{hyperref}

\begin{document}

\preprint{Draft --- not for distribution}

%
% The title and the list of authors
%
\title{Phonon anomalies in some iron-telluride materials}
\author{C. C. Homes}
\email{homes@bnl.gov}
\affiliation{Condensed Matter Physics and Materials Science Department,
  Brookhaven National Laboratory, Upton, New York 11973, USA}%
\author{Y. M. Dai}
\email{ymdai@lanl.gov}
\altaffiliation{Los Alamos National Laboratory, Center for Integrated
  Nanotechnologies, MPA-CINT, MS K771, Los Alamos, New Mexico 87545, USA}
\author{J. Schneeloch}
\author{R. D. Zhong}
\author{G. D. Gu}
\affiliation{Condensed Matter Physics and Materials Science Department,
  Brookhaven National Laboratory, Upton, New York 11973, USA}%
\date{\today}

%
% The abstract goes here
%
\begin{abstract}
The detailed temperature dependence of the infrared-active mode in Fe$_{1.03}$Te ($T_N\simeq 68$~K)
and Fe$_{1.13}$Te ($T_N\simeq 56$~K) has been examined, and the position, width, strength, and
asymmetry parameter determined using an asymmetric Fano profile superimposed on an electronic background.
In both materials the frequency of the mode increases as the temperature is reduced; however, there is
also a slight asymmetry in the line shape, indicating that the mode is coupled to either spin or charge
excitations.  Below $T_N$ there is an anomalous decrease in frequency and the mode shows little temperature
dependence, at the same time becoming more symmetric, suggesting a reduction in spin- or electron-phonon
coupling.  The frequency of the infrared-active mode and the magnitude of the shift below $T_N$ are predicted
reasonably well by first-principles calculations; however, the predicted splitting of the mode is not observed.
In superconducting FeTe$_{0.55}$Se$_{0.45}$ ($T_c\simeq 14$~K) the infrared-active $E_u$ mode displays asymmetric
line shape at all temperatures, which is most pronounced between $100 - 200$~K, indicating the presence
of either spin- or electron-phonon coupling, which may be a necessary prerequisite for superconductivity
in this class of materials.
\end{abstract}
%
%  PACS numbers
%  63.20.-e     Phonons in crystal lattices
%  71.27.+a 	Strongly correlated electron systems; heavy fermions
%  78.30.-j     Infrared and Raman spectra
%
\maketitle

%
% The main body of the text
%
% Introduction
%
\section{Introduction}
The discovery of superconductivity in the iron-based materials has prompted a
thorough investigation of this class of materials in an effort to understand
the mechanism responsible for the superconductivity, as well as the normal
state from which it emerges \cite{johnston10}.  While much of this work has
focussed on the electronic and magnetic structure, lattice vibrations are
also useful for probing electron-phonon coupling and the effects of chemical
substitution \cite{xu15}.
Among the iron-based materials, iron telluride is of particular interest
because it is thought to be among the most strongly correlated of the
iron-chalcogenide materials \cite{yin11}.  At room temperature, the nearly
stoichiometric Fe$_{1+\delta}$Te is a tetragonal, paramagnetic metal that
undergoes a first-order structural and magnetic transition  to a monoclinic,
antiferromagnetic (AFM) metal at $T_{N}\simeq 68$~K \cite{bao09,han09,martinelli10,zaliznyak11,
zaliznyak12,fobes14}.  The introduction of excess iron leads to an increase in
the resistivity and the suppression of $T_N$ \cite{liu11,rodriguez11}.  The
substitution of Se for Te in FeTe$_{1-x}$Se$_{x}$ suppresses the  structural
and magnetic transition and results in a superconducting phase for a broad
range of compositions \cite{li09,liu10}, with a maximum critical temperature of
$T_c \simeq 14$~K \cite{fang08,taen09,sales09,chen09}.
The optical and transport properties of Fe$_{1+\delta}$Te and FeTe$_{1-x}$Se$_x$
have been investigated \cite{chen09,hancock10,dai14,homes10,homes15,pimenov13,
perucchi14} and in addition there have been a number of reports on the
vibrational properties of these materials; however, only the Raman-active modes have
been studied \cite{xia09,okazaki11,gnezdilov11,um12,popovic14}, leaving the infrared-active modes
largely unexplored.

%
% This work
%
In this work we examine the optical properties of Fe$_{1.03}$Te ($T_N \simeq 68\,{\rm K}$)
and Fe$_{1.13}$Te ($T_N \simeq 56\,{\rm K}$) above and below $T_N$, as well as FeTe$_{0.55}$Se$_{0.45}$
($T_c\simeq 14$~K), and determine the detailed temperature dependence of the position,
width, strength and asymmetry parameter of the infrared-active mode using an asymmetric (Fano)
line shape; in several cases the electronic properties of these materials have been previously
discussed by us \cite{homes10,dai14,homes15}.  In both Fe$_{1+\delta}$Te materials the
infrared-active mode displays a slight asymmetry, suggesting either spin- or electron-phonon
coupling, and increases in frequency with decreasing temperature, as expected
for an anharmonic decay process; below $T_N$ the mode undergoes an anomalous decrease in
frequency at $T_N$ and the asymmetry parameter decreases, indicating reduced
coupling.
The vibrational frequencies and atomic intensities have been calculated at the center of
the Brillouin zone from first-principles methods for both the high-temperature tetragonal phase,
and the low-temperature monoclinic phase.  In the latter case, spin ordering is shown in
some cases to alter the character of the mode resulting in a large predicted frequency shift
below $T_N$.  Interestingly, the predicted splitting of the infrared-active $E_u$ mode below
$T_N$ is not observed.
In FeTe$_{0.55}$Se$_{0.45}$ the $E_u$ mode displays an asymmetry at all temperatures, which
is most pronounced between $100 - 200$~K.  The asymmetric profile is a signature of either spin-
or electron-phonon coupling, which may be necessary condition for superconductivity in this class
of materials.

%
% Experiment
%
\section{Experiment}
Single crystals of Fe$_{1.03}$Te and Fe$_{1.13}$Te with good cleavage planes
(001) have been grown by a unidirectional solidification process; these
crystals undergo structural and magnetic transitions \cite{bao09,han09,martinelli10,zaliznyak11,
zaliznyak12,fobes14} at $T_N \simeq 68$~K and $\simeq 56$~K, respectively.  Single
crystals with a nominal composition of FeTe$_{0.55}$Se$_{0.45}$ have also been
grown using this similar process with $T_c \simeq 14$~K and a transition width
of $\simeq 1$~K.  The reflectance of these materials with the light polarized in the
Fe--As (\emph{a-b}) planes has been measured over a wide frequency range
($\simeq 2$~meV to over 3~eV) using an \emph{in situ} evaporation
technique \cite{homes93} for a wide range of temperatures.  The complex optical
properties have been calculated from a Kramers-Kronig analysis of the
reflectance \cite{dressel-book}.  The large-scale structure and temperature dependence
of the optical conductivity of some of these materials have been previously
discussed \cite{homes10,dai14,homes15}.

%
% Results and discussion
%
\section{Results and Discussion}

Lattice vibrations in solids are often described using a Lorentzian oscillator
superimposed on a linear (or polynomial) background.  The complex dielectric function
$\tilde\epsilon= \epsilon_1+i\epsilon_2$ for the Lorentz oscillator is
\begin{equation}
  \tilde\epsilon(\omega) = \frac{\Omega_0^2}{\omega_0^2 - \omega^2 - i\gamma_0\omega},
\end{equation}
where $\omega_0$, $\gamma_0$ and $\Omega_0$ are the position, width, and strength of the
vibration, respectively.  The complex conductivity is $\tilde\sigma(\omega) = \sigma_1 +i\sigma_2 =
-2\pi i \,\omega [\tilde\epsilon(\omega) - \epsilon_\infty ]/Z_0$, where $\epsilon_\infty$
is a high-frequency contribution to the real part, and $Z_0 \simeq 377$~$\Omega$ is the
impedance of free space, yielding units for the conductivity of $\Omega^{-1}$cm$^{-1}$.
The real part of the optical conductivity for the oscillator may then be written as
\begin{equation}
  \sigma_1(\omega) = \frac{2\pi}{Z_0} \left[
    \frac{\gamma_0\,\omega^2\Omega_0^2}{(\omega_0^2-\omega_2)^2+\gamma_0^2\omega^2} \right].
   \label{eq:loren}
\end{equation}
While this approach accurately describes a symmetric line shape,  the coupling of lattice
vibrations to either the spin or electronic background may lead to an asymmetric line shape,
often referred to as the Fano profile \cite{fano61}.  The Fano line shape is written as
$\sigma_1(\omega) = A\left[(x+q)^2/(1+x^2)\right]$, where A is a constant,
$x=(\omega-\omega_0)/\gamma_0$, and the asymmetry is described by the dimensionless
parameter $1/q^2$.  A more useful expression,
\begin{equation}
  \sigma_1(\omega) = \frac{2\pi}{Z_0} \frac{\Omega_0^2}{\gamma_0}
  \frac{(q^2+4qx-1)}{q^2(1+4x^2)},
\end{equation}
incorporates the oscillator strength; however, this from contains no information
about $\sigma_2(\omega)$ and does not satisfy the Kramers-Kronig relation.
To resolve this issue a phenomenological dielectric function for a Fano-shaped
Lorentz oscillator is employed in which is Kramers-Kronig consistent
\cite{damascelli96,reffit},
\begin{equation}
  \tilde\epsilon(\omega) = \frac{\Omega_0^2}{\omega_0^2-\omega^2-i\gamma_0\omega}
  \left( 1+i\frac{\omega_q}{\omega} \right)^2 +
  \left( \frac{\Omega_0\omega_q}{\omega_0\omega} \right)^2
  \label{eq:fano}
\end{equation}
where $1/q=\omega_q/\omega_0$.  Unlike the simple Fano profile, a complex conductivity
may be calculated which satisfies $\tilde\sigma^\ast(\omega)=\tilde\sigma(-\omega)$.
The real part of the optical conductivity is then
%
% sigma1
%
\begin{equation}
  \sigma_1(\omega) = \frac{2\pi}{Z_0}
  \frac{\Omega_0^2\left[ \gamma_0\omega^2 - 2(\omega^2\omega_0-\omega_0^3)/q - \gamma_0\omega_0^2/q^2\right]}
   {(\omega^2 - \omega_0^2)^2+\gamma_0^2 \omega^2}.
%
%  \frac{\Omega_0^2\left[ q^2\gamma_0\omega^2 - 2q(\omega^2\omega_0-\omega_0^3) - \gamma_0\omega_0^2\right]}
%   {q^2 \left[\gamma_0^2 \omega^2 + (\omega^2 - \omega_0^2)^2\right]}.
\end{equation}
%
%and
%
% sigma2
%
%\begin{equation}
%  \sigma_2(\omega) = \frac{2\pi}{Z_0}
%  \frac{\omega\Omega_0^2\left[ (\omega^2-\omega_0^2)-2\gamma_0\omega_0/q +(\omega^2-\omega_0^2+\gamma_0^2)/q^2\right]}
%   {(\omega^2 - \omega_0^2)^2+\gamma_0^2 \omega^2}.
%\end{equation}
%
In the $1/q^2\rightarrow 0$ limit a symmetric Lorentzian is recovered [Eq.~(\ref{eq:loren})];
however, as $1/q^2$ increases the line shape becomes increasingly asymmetric.  We adopt this
latter form for asymmetric line shapes to fit the infrared-active vibrations in Fe$_{1+\delta}$Te
and FeTe$_{0.55}$Se$_{0.45}$.

%
% FeTe - 1.03 and 1.13
%
\subsection{\boldmath Fe$_{1+\delta}$Te \unboldmath}
\subsubsection{Vibrational properties}

%
% Figure 1: sigma_1 and fit at 30 K in Fe1.03Te
%
\begin{figure}[t]
%
% manuscript
%
%\vspace*{-0.5cm}%
\centerline{\includegraphics[width=3.4in]{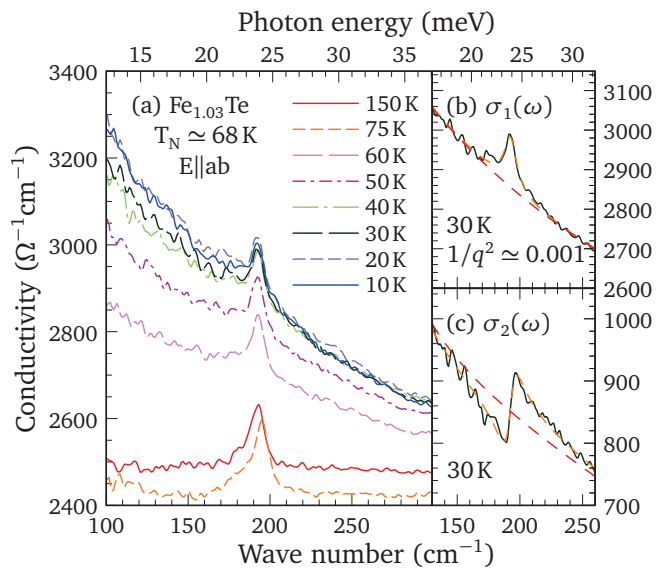}}%
%\vspace*{-1.3cm}%
\caption{(Color online) (a) The real part of the optical conductivity of
Fe$_{1.03}$Te above and below $T_N\simeq 68$~K in the region of the
infrared-active mode at $\simeq 196$~cm$^{-1}$.  There is no
evidence that this mode splits below $T_N$.
(b) Fano oscillator fit (dashed line) to $\sigma_1(\omega)$ at 30~K
with Drude-Lorentz background (long-dashed line), and (c) the simultaneous
fit to $\sigma_2(\omega)$; note the almost perfectly symmetric profile.}
%
%\vspace*{-0.0cm}%
\label{fig:sigma}
\end{figure}

The real part of the optical conductivity of Fe$_{1.03}$Te in the region of the
infrared-active phonon is shown in Fig.~\ref{fig:sigma}(a).  The rapid increase in
the low-frequency conductivity below $T_N$ has been attributed to the closing
of the pseudogap below $T_N$ leading to an increase in the free-carrier
concentration \cite{dai14}.
In the room-temperature tetragonal phase, the $P4/nmm$ space group with two
formula units per unit cell yields the irreducible vibrational representation \cite{xia09}
$$
  \Gamma_{\rm HT}=A_{1g}+B_{1g}+2E_g+A_{2u}+E_u.
$$
Two infrared-active modes are expected, a doubly-degenerate $E_u$ mode
active in the planes, and an $A_{2u}$ mode active along the \emph{c} axis.
In the low-temperature monoclinic phase, the $P2_1/m$ space group with
two formula units per unit cell yields the irreducible vibrational representation \cite{gnezdilov11}
$$
  \Gamma_{\rm LT}=4A_g+2B_g+A_u+2B_u,
$$
where the $A_u$ mode is infrared active along the \emph{c} axis and the $E_u$
modes splits into two $B_u$ modes that are active in the \emph{a-b} plane.

%
% Figure 2: phonon parameters for Fe1.03Te
%
\begin{figure}[b]
%
% manuscript
%
%\vspace*{-0.5cm}%
\centerline{\includegraphics[width=3.4in]{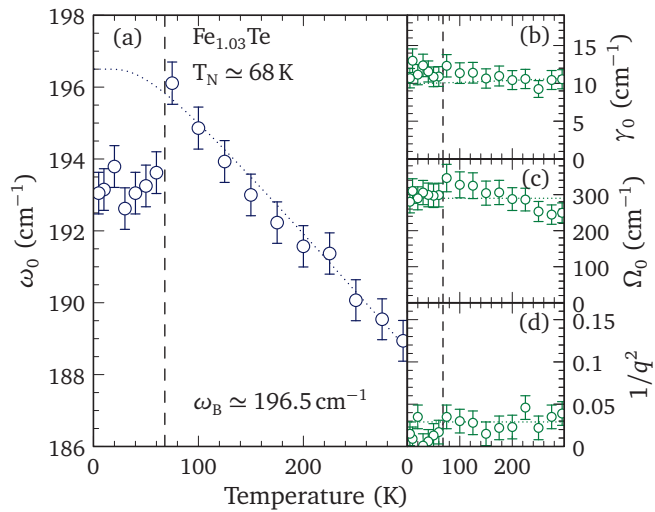}}%
%\vspace*{-1.3cm}%
\caption{(Color online) The temperature dependence of the
(a) infrared-active mode compared with the anharmonic
decay model (dotted line), (b) linewidth and predicted
dependence (dotted line), (c) oscillator strength, and
(d) asymmetry parameter in Fe$_{1.03}$Te ($T_N\simeq 68$~K);
the dotted lines in the last two panels are drawn as a
guide to the eye.}
%\vspace*{-0.0cm}%
\label{fig:ft103}
\end{figure}

The infrared-active mode is fit simultaneously to the real and imaginary
parts of the complex conductivity using the asymmetric line shape described by
the dielectric function in Eq.~(\ref{eq:fano}) superimposed on a free-carrier
(Drude) response in combination with several symmetric Lorentz oscillators that
describe the mid-infrared response.  Because the iron-telluride compounds are
multiband materials \cite{subedi08} the minimal description of the dielectric
function usually consists of at least two Drude components \cite{wu10a},
\begin{equation}
  \tilde\epsilon(\omega) = \epsilon_\infty - \sum_{j=1}^2
  \frac{\omega_{p,D;j}^2}{\omega^2+i\omega/\tau_{D,j}}
\end{equation}
where $\epsilon_\infty$ has been previously defined, $\omega_{p,D;j}^2 =
4\pi n_j e^2/m_j^\ast$ and $1/\tau_{D,j}$ are the square of the plasma frequency and
scattering rate for the delocalized (Drude) carriers, respectively, in the $j$th
band, and $n_j$ and $m_j^\ast$ are the carrier concentration and effective mass.
The values for $\omega_{p,D;j}$ and $1/\tau_{D,j}$ and the mid-infrared oscillators
are initially fit to the complex conductivity up to at least 1~eV using a non-linear
least-squares technique.  Fits to the infrared-active mode encompass only a narrow region
around the mode (typically $\pm 50$~cm$^{-1}$), and as a result in addition to the
vibrational parameters, only the Drude components and the width and strength of the
lowest mid-infrared oscillator are allowed to vary, while the other parameters remain
fixed.  The temperature dependence of $\omega_{p,D;1}$, $1/\tau_{D,j}$ and the mid-infrared
oscillators in these materials have been previously discussed \cite{dai14,homes15}.
As Figs.~\ref{fig:sigma}(b) and \ref{fig:sigma}(c) illustrate, both the background and the
line shape are reproduced quite well using this approach.

%
% Figure 3
%
\begin{figure}[t]
%
% manuscript
%
%\vspace*{-0.5cm}%
\centerline{\includegraphics[width=3.4in]{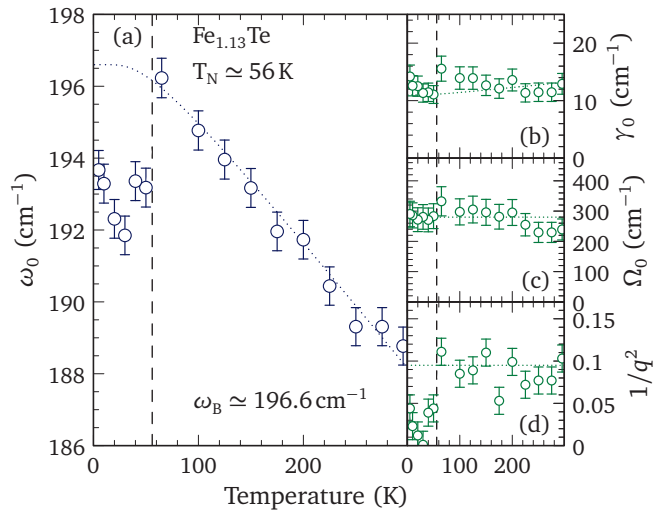}}%
%\vspace*{-1.3cm}%
\caption{(Color online) The temperature dependence of the
(a) infrared-active mode compared with the anharmonic
decay model (dotted line), (b) linewidth and predicted
dependence (dotted line), (c) oscillator strength, and
(d) asymmetry parameter in Fe$_{1.13}$Te ($T_N\simeq 56$~K);
the dotted lines in the last two panels are drawn as a guide
to the eye.}
%\vspace*{-0.0cm}%
\label{fig:ft113}
\end{figure}
%

%
% Table I
%
\begin{table*}[t]
\caption{Calculated frequencies and Fe atomic intensities in the high- and low-temperature phases of FeTe for the
zone center modes.  For each phase, the results based on the relaxed and experimental (shown in parenthesis) unit
cell parameters are considered; for the low-temperature phase, non-spin polarized (NSP) and spin polarized (SP)
calculations are performed.  All units are in cm$^{-1}$ unless otherwise indicated.}
\begin{ruledtabular}
\begin{tabular}{ccc c ccc cc c c}
    \multicolumn{2}{c}{$P4/nmm$ (HT)} & Atomic intensity & &
    \multicolumn{2}{c}{$P2_1/m$ (LT)} & Atomic intensity & & Atomic intensity & & \\

    Symmetry              & Theory                      & Fe              & &
    Symmetry & Theory-NSP & Fe    &  Theory-SP  & Fe & & Expt.$^a$  \\
    \cline{1-3} \cline{5-9} \cline{11-11}
% Eg -> Ag, Bg
    \multirow{2}{*}{$E_g$} & \multirow{2}{*}{98 (77)}   & \multirow{2}{*}{0.11 (0.12)} & \multirow{2}{*}{\large$\{$} & $A_g$ &  98\ (57)  & 0.10 (0.17) &  84  (81) & 0.13  (0.11) & & \\
                           &                            &                              & & $B_g$ &  99\ (72)  & 0.10 (0.13) &  85  (79) & 0.13  (0.08) & & \\
% A1g -> Ag
    $A_{1g}$               & 170 (141)                  &        0.00 (0.00)           & & $A_g$ & 173\ (137) & 0.00 (0.00) & 145  (139) & 0.15 (0.00) & & 160 \\
% Eu -> 2Bu
    \multirow{2}{*}{$E_u$} & \multirow{2}{*}{249 (203)} & \multirow{2}{*}{0.70 (0.70)} & \multirow{2}{*}{\large$\{$} & $B_u$ & 245\ (203) & 0.70 (0.70) & 241  (195) & 0.70 (0.70) & & \multirow{2}{*}{196} \\
                           &                            &                              & & $B_u$ & 300\ (254) & 0.70 (0.70) & 306  (267) & 0.70 (0.70) & & \\
% B1g -> Ag
    $B_{1g}$               & 272 (221)                  &        1.00 (1.00)           & & $A_g$ & 270\ (210) & 1.00 (0.89) & 167  (\,---\,) & 0.85 (\,\,---\,\,)  & & 203 \\
% Eg -> Ag, Bg
    \multirow{2}{*}{$E_g$} & \multirow{2}{*}{292 (223)} & \multirow{2}{*}{0.89 (0.88)} & \multirow{2}{*}{\large$\{$}& $A_g$ & 289\ (226) & 0.90 (0.92) & 287  (243) & 0.87 (0.89) & & \\
                           &                            &                              & & $B_g$ & 292\ (215) & 0.90 (0.87) & 288  (245) & 0.87 (0.92) & & \\
    $A_{2u}$               & 304 (257)                  &        0.70 (0.69)           & & $A_u$ & 248\ (196) & 0.70 (0.70) & 244  (189) & 0.70 (0.70) & & \\
\end{tabular}
\end{ruledtabular}
\footnotetext[1] {The Raman results are taken from Ref.~\onlinecite{um12}; infrared is from this work.}
\label{tab:freqs}
\end{table*}
%

%
% Fe1.03Te
%
The temperature dependence of $\omega_0$, $\gamma_0$, $\Omega_0$, and the asymmetry
parameter $1/q^2$ of the infrared-active mode in Fe$_{1.03}$Te are shown in Fig.~\ref{fig:ft103}.
The frequency of the mode increases (hardens) in an almost linear fashion
with decreasing temperature.  At $T_N$ there is an anomalous decrease (softening) in the
frequency of the mode followed by a very weak temperature dependence; there is no evidence
for a new infrared active $B_u$ mode appearing below $T_N$.  This is similar to the temperature
dependence of the frequency of the $B_{1g}$ mode in these materials \cite{gnezdilov11,um12,popovic14}.
The width of this mode shows very little temperature dependence either above or below
$T_N$.  While strength of this mode may increase slightly as the temperature is lowered
to $T_N$, it is also possible that, within error, it is constant; below $T_N$ there
is very little change.  Above $T_N$, the asymmetry parameter $1/q^2\simeq 0.03$,
suggesting weak spin- or electron-phonon coupling \cite{xu15}.  Below $T_N$ the asymmetry
parameter $1/q^2$ decreases and the line shape becomes almost perfectly symmetric; this
inferred reduction in the electron-phonon coupling may be related to the fact that
in the AFM state the spin fluctuations are observed to decrease \cite{zhang10} and
most signs of correlations disappear \cite{lin13}.

%
% Fe1.13Te
%
The temperature dependence of the infrared-active mode in Fe$_{1.13}$Te has been fit
using procedure described above, and the results are shown in Fig.~\ref{fig:ft113}
(the temperature dependence of the optical conductivity and the fit at 30~K
is shown in Fig.~S1 in the Supplemental Material).
The behavior of $\omega_0$, $\gamma_0$, and $\Omega_0$ are quite similar to those of
Fe$_{1.03}$Te, although there does tend to be a bit more scatter in the data points.
However, above $T_N$ the asymmetry parameter $1/q^2\simeq 0.1$, suggesting that
the spin- or electron-phonon coupling is stronger in this material; as in
Fe$_{1.03}$Te, below $T_N$ the asymmetry parameter $1/q^2$ decreases, suggesting a
decrease in correlations and a commensurate reduction in the coupling of the
phonons to either the spin or charge.

Above $T_N$ the behavior of $\omega_0$ can be described quite well assuming a
symmetric anharmonic decay into two acoustic modes with identical frequencies and
opposite momenta \cite{klemens66,menendez84}.  A slightly modified functional
form is employed here,
\begin{equation}
  \omega_{0}(T)=\omega_{\rm B} \left[1-\frac{2{\rm C}}{e^x-1}  \right],
\end{equation}
\begin{equation}
  \gamma_{0}(T)=\Gamma_{0} \left[1+\frac{2\Gamma}{e^x-1} \right],
\end{equation}
where $\omega_{\rm B}$ is the bare phonon frequency, $\Gamma_{0}$ is a residual line
width, C and $\Gamma$ are constants, and $x=\hbar\omega_{\rm B}/(2k_{\rm B}T)$;
an advantage of this form is that the bare phonon frequency (residual line
width) is recovered in the $T\rightarrow 0$ limit.  The model fits to the
$T>T_N$ data are indicated by the dotted lines in Figs.~\ref{fig:ft103} and
\ref{fig:ft113}, yielding nearly identical values of $\omega_B \simeq
196.5$ and 196.6~cm$^{-1}$ for Fe$_{1.03}$Te and Fe$_{1.13}$Te, respectively.

The behavior of the infrared-active mode at $T_N$ is also reminiscent of the
$E_u$ mode in BaFe$_2$As$_2$ which also experiences an anomalous softening below
the magnetic and structural transition at $T_N \simeq 138$~K \cite{akrap09};
however, in that material the oscillator strength nearly doubles below $T_N$,
whereas in Fe$_{1+\delta}$Te it remains more or less unchanged.

%
% Lattice dynamics
%
\subsubsection{Lattice dynamics}
%
% Phonons for high-temperature phase
%
The frequencies and atomic intensities (square of the vibrational amplitude of
each atom that sum to unity) for the zone-center phonons have been calculated
from first principles using the so-called frozen-phonon method, the details of
which are described the Appendix.
The results for the high-temperature tetragonal phase using the relaxed
(experimental) unit cell parameters (Table~\ref{tab:feteht} in the Appendix)
are shown in Table~\ref{tab:freqs} and compared with experimental results;
these results are in good agreement with other calculations \cite{xia09,okazaki11}.
It has been remarked that in this material the frequencies calculated using the
experimental unit cell parameters are in better agreement with the experimentally-observed
frequencies than the values determined from a relaxed unit cell \cite{xia09},
and that is indeed the case here.  In the relaxed unit cell the Fe--Te bond is
shorter, which has the effect of scaling up all of the frequencies by $20-30$\%;
however, we note that the character of the atomic intensities is essentially
unaffected.  The $A_{1g}$ and $B_{1g}$ Raman modes are calculated using the relaxed
(experimental) unit cell parameters at 170 (141) and 272 (221)~cm$^{-1}$, respectively,
are pure Te and Fe vibrations, whereas the $E_u$ mode is a mixture of the two.
Using the experimental unit cell, the calculated frequency for the $E_u$ mode of 203~cm$^{-1}$
is quite close to the measured value of $\simeq 196$~cm$^{-1}$, as are the values
for the $A_{1g}$ and $B_{1g}$ modes \cite{um12}.

%
% Phonons for low-temperature phase: non-magnetic
%
The frequencies and atomic intensities for the low-temperature monoclinic phase
have also been calculated using the relaxed (experimental) unit cell parameters
(Table~\ref{tab:fetelt} in the Appendix), and the results listed in Table~\ref{tab:freqs}.
As previously noted, the lower symmetry results in the splitting of the doubly-degenerate
$E_g \rightarrow A_{g}+B_{g}$ and $E_u\rightarrow 2B_u$ modes.  In the
non-magnetic calculation (non-spin polarized), the calculated frequencies
for the relaxed and experimental unit cells match quite well with their
counterparts in the tetragonal phase, and the character of the modes does
not change substantially.  In the relaxed unit cell the splitting of the
upper and lower $E_g$ modes is negligible ($\simeq 2$~cm$^{-1}$); however
in the experimental unit cell this splitting is much larger, $\simeq 20$~cm$^{-1}$.
In both the experimental and relaxed geometries, the $E_u$ mode is predicted to
split by $\simeq 50$~cm$^{-1}$.  A simple empirical force-constant model \cite{vibratz}
reproduces the size of the splitting below $T_N$ and indicates that the two $B_u$ modes
should have roughly the same strength.  Using a splitting of $\simeq 50$~cm$^{-1}$, a
new mode might be expected at $\simeq 240$~cm$^{-1}$; however, there does not appear to
be any experimental evidence for a new mode below $T_N$ [Fig.~\ref{fig:sigma}(a) and
Fig.~S1(a) in the Supplementary Material].

%
% Magnetic phase
%
The non-spin polarized calculation does not take into consideration the antiferromagnetic
ground state of this material \cite{bao09}.  Accordingly, the spins on the two Fe
sites have been placed in an antiferromagnetic configuration while the Te atoms
remain non-magnetic; the frequencies and atomic characters are then calculated for
a spin-polarized case using both the relaxed and experimentally-determined unit cell
parameters, and the results listed in Table~\ref{tab:freqs}.  The introduction of
magnetic order has a significant effect on some modes changing both the frequency
and the character of the vibration; however, we note that the $E_u$, upper $E_g$
and $A_{2u}$ modes experience only a small effect.  The small predicted downward
shift in the lower $B_u$ mode of $\simeq 4-7$~cm$^{-1}$ for the spin-polarized case
is in good agreement with the experimentally-observed softening of this mode
of between $2-4$~cm$^{-1}$ in Fe$_{1+\delta}$Te below $T_N$.

%
% Figure 4
%
\begin{figure}[b]
%
% manuscript
%
%\vspace*{-0.5cm}%
\centerline{\includegraphics[width=3.4in]{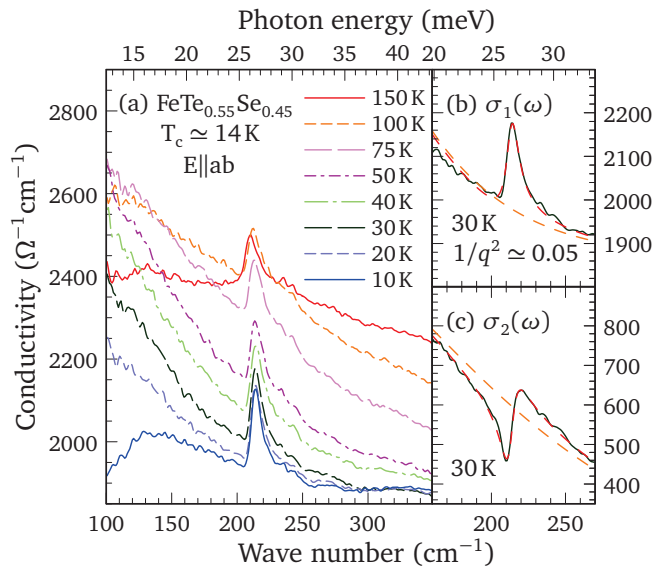}}%
%\vspace*{-1.3cm}%
\caption{(Color online) (a) The real part of the optical conductivity of
FeTe$_{0.55}$Se$_{0.45}$ above and below $T_c\simeq 14$~K in the region
of the infrared-active $E_u$ mode at $\simeq 213$~cm$^{-1}$.
(b) Fano oscillator fit (dashed line) to $\sigma_1(\omega)$  at 30~K
with Drude-Lorentz background (long-dashed line), and (c) the simultaneous
fit to $\sigma_2(\omega)$.}
%\vspace*{-0.0cm}%
\label{fig:fts}
\end{figure}

%
%%%%%%%%%%%%%%%%%%%%%%%%%%%%%%%%%%%%%%%%%%%%%%%%%%%%%%%%%%%%%%%%%%%%%%%
%
% FeTeSe
%
\subsection{\boldmath FeTe$_{0.55}$Se$_{0.45}$ \unboldmath}
The real part of the optical conductivity of superconducting FeTe$_{0.55}$Se$_{0.45}$
in the region of the infrared-active phonon at $\simeq 213$~cm$^{-1}$ is shown in
Fig.~\ref{fig:fts}(a) \cite{homes15}.  The substitution of Se for Te in this material
has the effect of shifting the frequency of the $E_u$ mode up about 10\%.  The temperature
dependence of the $E_u$ mode has been fit to the complex conductivity using the
asymmetric Fano line shape in Eq.~(\ref{eq:fano}) superimposed on a Drude-Lorentz
background; as Figs.~\ref{fig:fts}(b) and \ref{fig:fts}(c) illustrate, both the
background and the line shape are reproduced quite well.  In several cases, it appears
as thought there is a weak sideband at $\simeq 240$~cm$^{-1}$; however, no new mode
is predicted in this material, suggesting that such features are the result of an
asymmetric line shape.

%
% Figure 5
%
\begin{figure}[t]
%
% manuscript
%
%\vspace*{-0.5cm}%
\centerline{\includegraphics[width=3.4in]{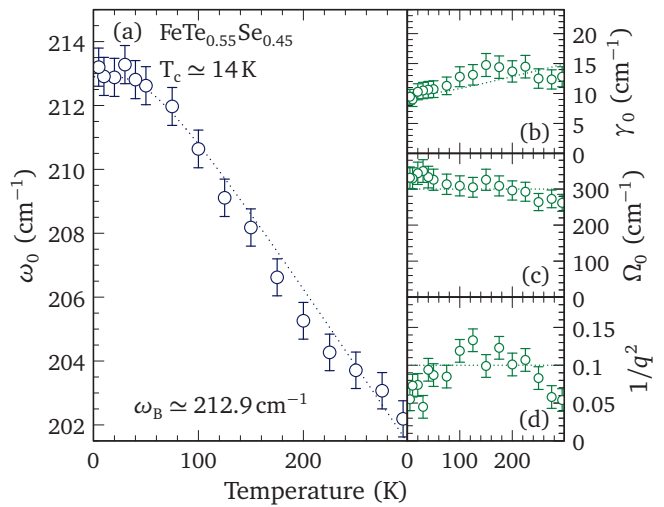}}%
%\vspace*{-1.3cm}%
\caption{(Color online) The temperature dependence of the
(a) infrared-active $E_u$ mode compared with the anharmonic
decay model (dotted line), (b) line width and predicted
dependence (dotted line), (c) oscillator strength, and
(d) asymmetry parameter in FeTe$_{0.55}$Se$_{0.45}$ ($T_c=14$~K);
the dotted lines in the last two panels are drawn as a guide to
the eye.}
%\vspace*{-0.0cm}%
\label{fig:fts_phon}
\end{figure}

The temperature dependence of the frequency and line width of the mode, shown in Figs.~\ref{fig:fts_phon}(a) and
\ref{fig:fts_phon}(b), respectively, both agree quite well with the anharmonic decay model.
The strength of the mode shown in Figs.~\ref{fig:fts_phon}(c) may be increasing somewhat
with decreasing temperature, but within the accuracy of the fits it is also possible that
it is temperature independent.
The asymmetry parameter is small at room temperature, $1/q^2\simeq 0.05$, but increases
with temperature, reaching a maximum of $1/q^2\simeq 0.13$ at $\simeq 125$~K before
decreasing at low temperature, as shown in
Fig.~\ref{fig:fts_phon}(d).  This would tend to indicate that the spin- or
electron-phonon coupling of the $E_u$ mode is becoming significant, and indeed
it is considerably larger than what is observed in related iron-pnictide
materials \cite{xu15}.
The decrease in $1/q^2$ at low temperature is correlated with the rapid decrease
in the Hall coefficient below 100~K \cite{tsukada10}, suggesting that the electron-phonon
coupling is related to a specific carrier pocket in this material.
While electron-phonon coupling is present in this material at low temperatures, it is
thought that the electron-phonon coupling constant $\lambda$ in this material is quite small
($\lambda \simeq 0.01$) \cite{luo12}.  The low values for $\lambda$ do not support the
high $T_{c}$'s observed in iron-based superconductors, implying that these materials are
not phonon-mediated superconductors \cite{boeri08}.  However, a detailed analysis of the
five-orbital Hubbard-Holstein model for iron pnictides has demonstrated that a relatively
small electron-phonon interaction arising from the Fe-ion oscillation can induce substantial
$d$-orbital fluctuations \cite{kotani10,saito10}, as well as the anion electronic polarization \cite{ma14}.
These orbital fluctuations give rise to a strong pairing interaction and mediate an $s_{++}$
superconducting state. In addition, a large Fe isotope effect has been reported in SmFeAsO$_{1-x}$Fe$_{x}$
and Ba$_{1-x}$K$_{x}$Fe$_{2}$As$_{2}$~\cite{liu09}, indicating that electron-phonon coupling likely
plays some role in the pairing mechanism. Our experimental results in the Fe$_{1+y}$Te$_{1-x}$Se$_{x}$
system, in combination with the above theoretical and experimental studies, suggests that electron-phonon
coupling may be a necessary prerequisite for superconductivity in this class of materials.

%
% Conclusions
%
\section{Conclusions}
The detailed temperature dependence of the infrared-active infrared-active mode has
been fit in Fe$_{1.03}$Te and Fe$_{1.13}$Te using an asymmetric Fano line shape
superimposed on a background of Drude free carriers and symmetric Lorentz oscillators.
In both materials, above $T_N$ the frequency increases as the temperature is
lowered; this behavior is described quite well by an anharmonic decay process.
Below $T_N$, there is an abrupt softening of this mode in both materials, below
which it displays little temperature dependence.  While frequency of the
mode and the size of the shift below $T_N$ are predicted reasonably well from
first-principles calculations (with some caveats), the predicted splitting of the
infrared-active mode is not observed.  In both materials, the phonon has a slightly
asymmetric line shape above $T_N$, suggesting that there is coupling of the phonons
to either spin or charge; however, $1/q^2$ decreases and the profile becomes more
symmetric below $T_N$, suggesting this coupling is reduced.
The $E_u$ mode in superconducting FeTe$_{0.55}$Se$_{0.45}$ has been fit in a similar
fashion and displays an asymmetry at all temperatures, which is most pronounced
between $100 - 200$~K.  The temperature dependence of this mode is once again
described quite well by an anharmonic decay process over the entire temperature
range.  The asymmetric nature of the $E_u$ mode is an indication of spin- or
electron-phonon coupling to the electronic background, which may be a necessary
prerequisite for superconductivity in the iron-based materials.

%
% acknowledgments
%
\vspace*{-2.0mm}
\begin{acknowledgements}
We would like to acknowledge helpful discussions with T. Birol.  We are
indebted to S. V. Dordevic for referring us to a phenomenological Fano line
shape, developed by A. Damascelli and A. Kuzmenko.
This work is supported by the Office of Science, U.S. Department of Energy
under Contract No. DE-SC0012704.  J.S. and R.D.Z. are supported by the Center
for Emergent Superconductivity, an Energy Frontier Research Center funded by
the U.S. Department of Energy, Office of Science.
\end{acknowledgements}

%
% Appendices
%
\appendix*
\section{Lattice dynamics}
%\section{Phonon calculations}
\subsection{\boldmath $P4/nmm$ \unboldmath (HT)}

The total energy of FeTe was calculated with the generalized gradient approximation
(GGA) using the full-potential linearized augmented plane-wave (FP-LAPW) method
\cite{singh} with local-orbital extensions \cite{singh91} in the WIEN2k implementation
\cite{wien2k}.  An examination of different Monkhorst-Pack {\em k}-point meshes
indicated that 175 $k$ points ($6\times{6}\times{4}$ mesh) and $R_{mt}k_{max}=7.5$
was sufficient for good energy convergence.  The geometry of the unit cell was refined
through an iterative process whereby the volume was optimized with respect to the total
energy while the $c/a$ ratio remained fixed.  The atomic fractional coordinates were then
relaxed with respect to the total force, typically resulting in residual forces of less
than 1~mRy/a.u.  This procedure was repeated until no further improvement was obtained.
A comparison of the experimental and calculated (relaxed) unit cell parameters are shown
in Table.~\ref{tab:feteht}.  Note that while the unit cell parameters are nearly identical
to the experimentally-determined values, the position of the Te atom has shifted somewhat.

%
% Table II
%
\begin{table}[t]
\caption{The experimental and theoretical lattice constants and atomic fractional
coordinates for the relaxed structure of FeTe for the high-temperature tetragonal
$P4/nmm$ space group.  The fractional coordinate for the Fe atom is
($\frac{3}{4} \frac{1}{4} 0$).  The Fe--Te bond lengths are also shown, while the Fe--Fe
and Te--Te bond lengths, 2.6960~\AA\ and 2.8126~\AA , respectively, are essentially
identical for the two structures.}
\begin{ruledtabular}
\begin{tabular}{ccc}
                                 & Experiment$^a$ & GGA \\
 $a$ (\AA )                      & 3.8127  & 3.8126  \\
 $c$ (\AA )                      & 6.2450  & 6.2448  \\
 Te($\frac{1}{4} \frac{1}{4} z$) & 0.2813 & 0.2569 \\
%
% Fe--Fe (\AA )                   & 2.6960 & 2.6959 \\
 Fe--Te (\AA )                   & 2.5923 & 2.4915 \\
% Te--Te (\AA )                   & 3.8127 & 2.8126 \\

\end{tabular}
\end{ruledtabular}
\footnotetext[1] {Ref~\onlinecite{bao09} [Supplemental Material. Table~I, Section (a)].}
\label{tab:feteht}
\end{table}

The phonons have been determined using the direct method.  To determine the
phonons at the zone center, a $1\times 1\times 1$ supercell is
sufficient.  To obtain a complete set of Hellmann-Feynman forces, a total of 4
independent displacements are required; because there are always some residual
forces at the atomic sites we have considered symmetric displacements,
which doubles this number, resulting in a total of 8 atomic displacements.
In this case, displacement amplitudes of 0.03~\AA\ were used.  The calculations
have converged when the successive changes for the forces on each atom are
less than 0.1~mRy/au.  The residual forces are collected for each set of
symmetric displacements and a list of the Hellmann-Feynman forces are generated.
Using the program PHONON \cite{phonon} the cumulative force constants deconvoluted
from the Hellmann-Feynman forces are introduced into the dynamical matrix, which
is then diagonalized in order to obtain the phonon frequencies; the atomic intensities
are further calculated to describe the character of the vibration.  The results are
shown in Table~\ref{tab:freqs}.

%
%
% Table III
%
\begin{table*}[htb]
\caption{The experimental and theoretical lattice constants and atomic fractional
coordinates for the relaxed structure of FeTe for the low-temperature monoclinic
$P2_1/m$ space group.  The Fe--Fe and Fe--Te bond lengths are also shown.  Note that
the GGA calculation does not reproduce the experimentally-observed Fe--Fe bond modulation.}
\begin{ruledtabular}
\begin{tabular}{c cc cc cc}
                                 & \multicolumn{2}{c}{Experiment$^a$} & \multicolumn{2}{c}{GGA (NSP)} & \multicolumn{2}{c}{GGA (SP)} \\
 $a$ (\AA )                      & \multicolumn{2}{c}{3.8346} & \multicolumn{2}{c}{3.8391} & \multicolumn{2}{c}{3.8391} \\
 $b$ (\AA )                      & \multicolumn{2}{c}{3.7836} & \multicolumn{2}{c}{3.7897} & \multicolumn{2}{c}{3.7897} \\
 $c$ (\AA )                      & \multicolumn{2}{c}{6.2567} & \multicolumn{2}{c}{6.2649} & \multicolumn{2}{c}{6.2649} \\
 $\beta$ ($^\circ$)              & \multicolumn{2}{c}{89.21}  & \multicolumn{2}{c}{89.24}  & \multicolumn{2}{c}{89.24} \\
 Fe($x\,\frac{1}{4}\,z$)         & 0.7612 & 0.0033            & 0.7501 & 0.0013 & 0.7507 & 0.0011 \\
 Te($x\,\frac{1}{4}\,z$)         & 0.2586 & 0.2822            & 0.2448 & 0.2559 & 0.2454 & 0.2717 \\
 Fe--Fe (\AA )                   & \multicolumn{2}{c}{2.633/2.756} & \multicolumn{2}{c}{2.697/2.698} & \multicolumn{2}{c}{2.693/2.701} \\
 Fe--Te (\AA )                   & \multicolumn{2}{c}{2.582/2.603/2.604} & \multicolumn{2}{c}{2.487/2.495/2.496} & \multicolumn{2}{c}{2.552/2.559/2.562}\\
% Te--Te                         &
\end{tabular}
\end{ruledtabular}
\footnotetext[1] {Ref~\onlinecite{zaliznyak12}.}
\label{tab:fetelt}
\end{table*}

%
% Low temperature phase
%
\subsection{\boldmath ${P2_1/m}$ \unboldmath (LT)}
The procedure that was used to calculate the geometry of the room-temperature
tetragonal phase has been repeated for the low-temperature monoclinic phase; the
same $k$ point mesh and $R_{mt}k_{max}$ were used.  The optimized unit cell
parameters are compared with the experimental values and shown in Table~\ref{tab:fetelt}.
Once again, the calculated unit cell parameters are nearly identical to the
experimental values; however the Fe and Te atoms have undergone small shifts
resulting in a decrease of the Fe--Te bond lengths.  Interestingly, the
Fe--Fe bond modulation observed experimentally is not captured in the non-spin
polarized GGA result in which the Fe--Fe bonds are almost identical.
To calculate the zone-center phonons, a $1\times 1\times 1$ supercell is
employed.  In the monoclinic phase a total of 6 independent displacements are
required; symmetric displacements double this number to 12.  Displacement
amplitudes of 0.03~\AA\ were used.

Two different cases have been considered for this symmetry.  In the first
case the system is assumed to be non-magnetic and the spins on the Fe
and Te sites are ignored and a non-spin polarized (NSP) calculation performed.
The calculated phonon frequencies and atomic intensities which are
listed in Table~\ref{tab:freqs}.
In the second case the spins on the Fe sites are specified to be in an
antiferromagnetic configuration and the Te atoms are assumed to be nonmagnetic
and a spin-polarized calculation is performed.  If the fractional coordinates are
relaxed in this case, position of the Te atom is now closer to the experimental
value, and a Fe--Fe bond modulation is now observed (Table~\ref{tab:fetelt}).
However, given that this structure appears to be intermediate between the
non-spin polarized and the experimental unit cell parameters, we have
not performed a phonon calculation.  Using the previously established criteria
for convergence, we find the magnetic moments are about $1.7\,\mu_B$/Fe for
the (non-spin polarized) relaxed unit cell, which is very close to the
expected value of $2\,\mu_B$/Fe observed in other work \cite{gretarsson11}.
When the experimental unit cell is used, the magnetic moments increase
to $2.2\,\mu_B$/Fe; however, this results in a calculated frequency of
less than zero for one of the $A_g$ modes, this unphysical value indicates
that within the scope of this calculation the structure is unstable.

%\vfill
%\eject

%
% acknowledgments
%
%We would like to acknowledge useful discussions with D. N. Basov, J. P.
%Carbotte, A. V. Chubukov and J. M. Tranquada.
%
%This work is supported by the Office of Science, U.S. Department of Energy (DOE)
%under Contract No. DE-AC02-98CH10886.

%
%%%%%%%%%%%%%%%%%%%%%%%%%%%%%%%%%%%%%%%%%%%%%%%%%%%%%%%%%%%%%%%%%%%%%%%%%%%%%%
%
% References
%
%\bibliography{phonons}

\begin{thebibliography}{53}%
\makeatletter
\providecommand \@ifxundefined [1]{%
 \@ifx{#1\undefined}
}%
\providecommand \@ifnum [1]{%
 \ifnum #1\expandafter \@firstoftwo
 \else \expandafter \@secondoftwo
 \fi
}%
\providecommand \@ifx [1]{%
 \ifx #1\expandafter \@firstoftwo
 \else \expandafter \@secondoftwo
 \fi
}%
\providecommand \natexlab [1]{#1}%
\providecommand \enquote  [1]{``#1''}%
\providecommand \bibnamefont  [1]{#1}%
\providecommand \bibfnamefont [1]{#1}%
\providecommand \citenamefont [1]{#1}%
\providecommand \href@noop [0]{\@secondoftwo}%
\providecommand \href [0]{\begingroup \@sanitize@url \@href}%
\providecommand \@href[1]{\@@startlink{#1}\@@href}%
\providecommand \@@href[1]{\endgroup#1\@@endlink}%
\providecommand \@sanitize@url [0]{\catcode `\\12\catcode `\$12\catcode
  `\&12\catcode `\#12\catcode `\^12\catcode `\_12\catcode `\%12\relax}%
\providecommand \@@startlink[1]{}%
\providecommand \@@endlink[0]{}%
\providecommand \url  [0]{\begingroup\@sanitize@url \@url }%
\providecommand \@url [1]{\endgroup\@href {#1}{\urlprefix }}%
\providecommand \urlprefix  [0]{URL }%
\providecommand \Eprint [0]{\href }%
\providecommand \doibase [0]{http://dx.doi.org/}%
\providecommand \selectlanguage [0]{\@gobble}%
\providecommand \bibinfo  [0]{\@secondoftwo}%
\providecommand \bibfield  [0]{\@secondoftwo}%
\providecommand \translation [1]{[#1]}%
\providecommand \BibitemOpen [0]{}%
\providecommand \bibitemStop [0]{}%
\providecommand \bibitemNoStop [0]{.\EOS\space}%
\providecommand \EOS [0]{\spacefactor3000\relax}%
\providecommand \BibitemShut  [1]{\csname bibitem#1\endcsname}%
\let\auto@bib@innerbib\@empty
%</preamble>
\bibitem [{\citenamefont {Johnston}(2010)}]{johnston10}%
  \BibitemOpen
  \bibfield  {author} {\bibinfo {author} {\bibfnamefont {David~C.}\
  \bibnamefont {Johnston}},\ }\bibfield  {title} {\enquote {\bibinfo {title}
  {The puzzle of high temperature superconductivity in layered iron pnictides
  and chalcogenides},}\ }\href {\doibase 10.1080/00018732.2010.513480}
  {\bibfield  {journal} {\bibinfo  {journal} {Adv. Phys.}\ }\textbf {\bibinfo
  {volume} {59}},\ \bibinfo {pages} {803--1061} (\bibinfo {year}
  {2010})}\BibitemShut {NoStop}%
\bibitem [{\citenamefont {Xu}\ \emph {et~al.}(2015)\citenamefont {Xu},
  \citenamefont {Dai}, \citenamefont {Shen}, \citenamefont {Xiao},
  \citenamefont {Ye}, \citenamefont {Forget}, \citenamefont {Colson},
  \citenamefont {Feng}, \citenamefont {Wen}, \citenamefont {Homes},
  \citenamefont {Qiu},\ and\ \citenamefont {Lobo}}]{xu15}%
  \BibitemOpen
  \bibfield  {author} {\bibinfo {author} {\bibfnamefont {B.}~\bibnamefont
  {Xu}}, \bibinfo {author} {\bibfnamefont {Y.~M.}\ \bibnamefont {Dai}},
  \bibinfo {author} {\bibfnamefont {B.}~\bibnamefont {Shen}}, \bibinfo {author}
  {\bibfnamefont {H.}~\bibnamefont {Xiao}}, \bibinfo {author} {\bibfnamefont
  {Z.~R.}\ \bibnamefont {Ye}}, \bibinfo {author} {\bibfnamefont
  {A.}~\bibnamefont {Forget}}, \bibinfo {author} {\bibfnamefont
  {D.}~\bibnamefont {Colson}}, \bibinfo {author} {\bibfnamefont {D.~L.}\
  \bibnamefont {Feng}}, \bibinfo {author} {\bibfnamefont {H.~H.}\ \bibnamefont
  {Wen}}, \bibinfo {author} {\bibfnamefont {C.~C.}\ \bibnamefont {Homes}},
  \bibinfo {author} {\bibfnamefont {X.~G.}\ \bibnamefont {Qiu}}, \ and\
  \bibinfo {author} {\bibfnamefont {R.~P. S.~M.}\ \bibnamefont {Lobo}},\
  }\bibfield  {title} {\enquote {\bibinfo {title} {{Anomalous phonon redshift
  in {K}-doped {BaFe}$_{2}${As}$_{2}$ iron pnictides}},}\ }\href {\doibase
  10.1103/PhysRevB.91.104510} {\bibfield  {journal} {\bibinfo  {journal} {Phys.
  Rev. B}\ }\textbf {\bibinfo {volume} {91}},\ \bibinfo {pages} {104510}
  (\bibinfo {year} {2015})}\BibitemShut {NoStop}%
\bibitem [{\citenamefont {Yin}\ \emph {et~al.}(2011)\citenamefont {Yin},
  \citenamefont {Haule},\ and\ \citenamefont {Kotliar}}]{yin11}%
  \BibitemOpen
  \bibfield  {author} {\bibinfo {author} {\bibfnamefont {Z.~P.}\ \bibnamefont
  {Yin}}, \bibinfo {author} {\bibfnamefont {K.}~\bibnamefont {Haule}}, \ and\
  \bibinfo {author} {\bibfnamefont {G.}~\bibnamefont {Kotliar}},\ }\bibfield
  {title} {\enquote {\bibinfo {title} {{Kinetic frustration and the nature of
  the magnetic and paramagnetic states in iron pnictides and
  iron chalcogenides}},}\ }\href {\doibase 10.1038/nmat3120} {\bibfield
  {journal} {\bibinfo  {journal} {Nat. Mater.}\ }\textbf {\bibinfo {volume}
  {10}},\ \bibinfo {pages} {932--935} (\bibinfo {year} {2011})}\BibitemShut
  {NoStop}%
\bibitem [{\citenamefont {Bao}\ \emph {et~al.}(2009)\citenamefont {Bao},
  \citenamefont {Qiu}, \citenamefont {Huang}, \citenamefont {Green},
  \citenamefont {Zajdel}, \citenamefont {Fitzsimmons}, \citenamefont
  {Zhernenkov}, \citenamefont {Chang}, \citenamefont {Fang}, \citenamefont
  {Qian}, \citenamefont {Vehstedt}, \citenamefont {Yang}, \citenamefont {Pham},
  \citenamefont {Spinu},\ and\ \citenamefont {Mao}}]{bao09}%
  \BibitemOpen
  \bibfield  {author} {\bibinfo {author} {\bibfnamefont {Wei}\ \bibnamefont
  {Bao}}, \bibinfo {author} {\bibfnamefont {Y.}~\bibnamefont {Qiu}}, \bibinfo
  {author} {\bibfnamefont {Q.}~\bibnamefont {Huang}}, \bibinfo {author}
  {\bibfnamefont {M.~A.}\ \bibnamefont {Green}}, \bibinfo {author}
  {\bibfnamefont {P.}~\bibnamefont {Zajdel}}, \bibinfo {author} {\bibfnamefont
  {M.~R.}\ \bibnamefont {Fitzsimmons}}, \bibinfo {author} {\bibfnamefont
  {M.}~\bibnamefont {Zhernenkov}}, \bibinfo {author} {\bibfnamefont
  {S.}~\bibnamefont {Chang}}, \bibinfo {author} {\bibfnamefont {Minghu}\
  \bibnamefont {Fang}}, \bibinfo {author} {\bibfnamefont {B.}~\bibnamefont
  {Qian}}, \bibinfo {author} {\bibfnamefont {E.~K.}\ \bibnamefont {Vehstedt}},
  \bibinfo {author} {\bibfnamefont {Jinhu}\ \bibnamefont {Yang}}, \bibinfo
  {author} {\bibfnamefont {H.~M.}\ \bibnamefont {Pham}}, \bibinfo {author}
  {\bibfnamefont {L.}~\bibnamefont {Spinu}}, \ and\ \bibinfo {author}
  {\bibfnamefont {Z.~Q.}\ \bibnamefont {Mao}},\ }\bibfield  {title} {\enquote
  {\bibinfo {title} {{Tunable ($\ensuremath{\delta}\ensuremath{\pi}$,
  $\ensuremath{\delta}\ensuremath{\pi}$)-Type Antiferromagnetic Order in
  $\ensuremath{\alpha}$-Fe(Te,Se) Superconductors}},}\ }\href {\doibase
  10.1103/PhysRevLett.102.247001} {\bibfield  {journal} {\bibinfo  {journal}
  {Phys. Rev. Lett.}\ }\textbf {\bibinfo {volume} {102}},\ \bibinfo {pages}
  {247001} (\bibinfo {year} {2009})}\BibitemShut {NoStop}%
\bibitem [{\citenamefont {Han}\ and\ \citenamefont {Savrasov}(2009)}]{han09}%
  \BibitemOpen
  \bibfield  {author} {\bibinfo {author} {\bibfnamefont {Myung~Joon}\
  \bibnamefont {Han}}\ and\ \bibinfo {author} {\bibfnamefont {Sergey~Y.}\
  \bibnamefont {Savrasov}},\ }\bibfield  {title} {\enquote {\bibinfo {title}
  {{Doping Driven ($\ensuremath{\pi}$,0) Nesting and Magnetic Properties of
  ${\mathrm{Fe}}_{1+x}\mathrm{Te}$ Superconductors}},}\ }\href {\doibase
  10.1103/PhysRevLett.103.067001} {\bibfield  {journal} {\bibinfo  {journal}
  {Phys. Rev. Lett.}\ }\textbf {\bibinfo {volume} {103}},\ \bibinfo {pages}
  {067001} (\bibinfo {year} {2009})}\BibitemShut {NoStop}%
\bibitem [{\citenamefont {Martinelli}\ \emph {et~al.}(2010)\citenamefont
  {Martinelli}, \citenamefont {Palenzona}, \citenamefont {Tropeano},
  \citenamefont {Ferdeghini}, \citenamefont {Putti}, \citenamefont {Cimberle},
  \citenamefont {Nguyen}, \citenamefont {Affronte},\ and\ \citenamefont
  {Ritter}}]{martinelli10}%
  \BibitemOpen
  \bibfield  {author} {\bibinfo {author} {\bibfnamefont {A.}~\bibnamefont
  {Martinelli}}, \bibinfo {author} {\bibfnamefont {A.}~\bibnamefont
  {Palenzona}}, \bibinfo {author} {\bibfnamefont {M.}~\bibnamefont {Tropeano}},
  \bibinfo {author} {\bibfnamefont {C.}~\bibnamefont {Ferdeghini}}, \bibinfo
  {author} {\bibfnamefont {M.}~\bibnamefont {Putti}}, \bibinfo {author}
  {\bibfnamefont {M.~R.}\ \bibnamefont {Cimberle}}, \bibinfo {author}
  {\bibfnamefont {T.~D.}\ \bibnamefont {Nguyen}}, \bibinfo {author}
  {\bibfnamefont {M.}~\bibnamefont {Affronte}}, \ and\ \bibinfo {author}
  {\bibfnamefont {C.}~\bibnamefont {Ritter}},\ }\bibfield  {title} {\enquote
  {\bibinfo {title} {{From antiferromagnetism to superconductivity in
  ${\text{Fe}}_{1+y}{\text{Te}}_{1\ensuremath{-}x}{\text{Se}}_{x}$
  $(0\ensuremath{\le}x\ensuremath{\le}0.20)$: Neutron powder diffraction
  analysis}},}\ }\href {\doibase 10.1103/PhysRevB.81.094115} {\bibfield
  {journal} {\bibinfo  {journal} {Phys. Rev. B}\ }\textbf {\bibinfo {volume}
  {81}},\ \bibinfo {pages} {094115} (\bibinfo {year} {2010})}\BibitemShut
  {NoStop}%
\bibitem [{\citenamefont {Zaliznyak}\ \emph {et~al.}(2011)\citenamefont
  {Zaliznyak}, \citenamefont {Xu}, \citenamefont {Tranquada}, \citenamefont
  {Gu}, \citenamefont {Tsvelik},\ and\ \citenamefont {Stone}}]{zaliznyak11}%
  \BibitemOpen
  \bibfield  {author} {\bibinfo {author} {\bibfnamefont {Igor~A.}\ \bibnamefont
  {Zaliznyak}}, \bibinfo {author} {\bibfnamefont {Zhijun}\ \bibnamefont {Xu}},
  \bibinfo {author} {\bibfnamefont {John~M.}\ \bibnamefont {Tranquada}},
  \bibinfo {author} {\bibfnamefont {Genda}\ \bibnamefont {Gu}}, \bibinfo
  {author} {\bibfnamefont {Alexei~M.}\ \bibnamefont {Tsvelik}}, \ and\ \bibinfo
  {author} {\bibfnamefont {Matthew~B.}\ \bibnamefont {Stone}},\ }\bibfield
  {title} {\enquote {\bibinfo {title} {{Unconventional Temperature Enhanced
  Magnetism in {Fe}$_{1.1}${Te} }},}\ }\href {\doibase
  10.1103/PhysRevLett.107.216403} {\bibfield  {journal} {\bibinfo  {journal}
  {Phys. Rev. Lett.}\ }\textbf {\bibinfo {volume} {107}},\ \bibinfo {pages}
  {216403} (\bibinfo {year} {2011})}\BibitemShut {NoStop}%
\bibitem [{\citenamefont {Zaliznyak}\ \emph {et~al.}(2012)\citenamefont
  {Zaliznyak}, \citenamefont {Xu}, \citenamefont {Wen}, \citenamefont
  {Tranquada}, \citenamefont {Gu}, \citenamefont {Solovyov}, \citenamefont
  {Glazkov}, \citenamefont {Zheludev}, \citenamefont {Garlea},\ and\
  \citenamefont {Stone}}]{zaliznyak12}%
  \BibitemOpen
  \bibfield  {author} {\bibinfo {author} {\bibfnamefont {I.~A.}\ \bibnamefont
  {Zaliznyak}}, \bibinfo {author} {\bibfnamefont {Z.~J.}\ \bibnamefont {Xu}},
  \bibinfo {author} {\bibfnamefont {J.~S.}\ \bibnamefont {Wen}}, \bibinfo
  {author} {\bibfnamefont {J.~M.}\ \bibnamefont {Tranquada}}, \bibinfo {author}
  {\bibfnamefont {G.~D.}\ \bibnamefont {Gu}}, \bibinfo {author} {\bibfnamefont
  {V.}~\bibnamefont {Solovyov}}, \bibinfo {author} {\bibfnamefont {V.~N.}\
  \bibnamefont {Glazkov}}, \bibinfo {author} {\bibfnamefont {A.~I.}\
  \bibnamefont {Zheludev}}, \bibinfo {author} {\bibfnamefont {V.~O.}\
  \bibnamefont {Garlea}}, \ and\ \bibinfo {author} {\bibfnamefont {M.~B.}\
  \bibnamefont {Stone}},\ }\bibfield  {title} {\enquote {\bibinfo {title}
  {{Continuous magnetic and structural phase transitions in Fe$_{1+y}$Te}},}\
  }\href {\doibase 10.1103/PhysRevB.85.085105} {\bibfield  {journal} {\bibinfo
  {journal} {Phys. Rev. B}\ }\textbf {\bibinfo {volume} {85}},\ \bibinfo
  {pages} {085105} (\bibinfo {year} {2012})}\BibitemShut {NoStop}%
\bibitem [{\citenamefont {Fobes}\ \emph {et~al.}(2014)\citenamefont {Fobes},
  \citenamefont {Zaliznyak}, \citenamefont {Xu}, \citenamefont {Zhong},
  \citenamefont {Gu}, \citenamefont {Tranquada}, \citenamefont {Harriger},
  \citenamefont {Singh}, \citenamefont {Garlea}, \citenamefont {Lumsden},\ and\
  \citenamefont {Winn}}]{fobes14}%
  \BibitemOpen
  \bibfield  {author} {\bibinfo {author} {\bibfnamefont {David}\ \bibnamefont
  {Fobes}}, \bibinfo {author} {\bibfnamefont {Igor~A.}\ \bibnamefont
  {Zaliznyak}}, \bibinfo {author} {\bibfnamefont {Zhijun}\ \bibnamefont {Xu}},
  \bibinfo {author} {\bibfnamefont {Ruidan}\ \bibnamefont {Zhong}}, \bibinfo
  {author} {\bibfnamefont {Genda}\ \bibnamefont {Gu}}, \bibinfo {author}
  {\bibfnamefont {John~M.}\ \bibnamefont {Tranquada}}, \bibinfo {author}
  {\bibfnamefont {Leland}\ \bibnamefont {Harriger}}, \bibinfo {author}
  {\bibfnamefont {Deepak}\ \bibnamefont {Singh}}, \bibinfo {author}
  {\bibfnamefont {V.~Ovidiu}\ \bibnamefont {Garlea}}, \bibinfo {author}
  {\bibfnamefont {Mark}\ \bibnamefont {Lumsden}}, \ and\ \bibinfo {author}
  {\bibfnamefont {Barry}\ \bibnamefont {Winn}},\ }\bibfield  {title} {\enquote
  {\bibinfo {title} {Ferro-orbital ordering transition in iron telluride
  {Fe}$_{1+y}${Te}},}\ }\href {\doibase 10.1103/PhysRevLett.112.187202}
  {\bibfield  {journal} {\bibinfo  {journal} {Phys. Rev. Lett.}\ }\textbf
  {\bibinfo {volume} {112}},\ \bibinfo {pages} {187202} (\bibinfo {year}
  {2014})}\BibitemShut {NoStop}%
\bibitem [{\citenamefont {Liu}\ \emph {et~al.}(2011)\citenamefont {Liu},
  \citenamefont {Lee}, \citenamefont {Xu}, \citenamefont {Wen}, \citenamefont
  {Gu}, \citenamefont {Ku}, \citenamefont {Tranquada},\ and\ \citenamefont
  {Hill}}]{liu11}%
  \BibitemOpen
  \bibfield  {author} {\bibinfo {author} {\bibfnamefont {X.}~\bibnamefont
  {Liu}}, \bibinfo {author} {\bibfnamefont {C.-C.}\ \bibnamefont {Lee}},
  \bibinfo {author} {\bibfnamefont {Z.~J.}\ \bibnamefont {Xu}}, \bibinfo
  {author} {\bibfnamefont {J.~S.}\ \bibnamefont {Wen}}, \bibinfo {author}
  {\bibfnamefont {G.}~\bibnamefont {Gu}}, \bibinfo {author} {\bibfnamefont
  {W.}~\bibnamefont {Ku}}, \bibinfo {author} {\bibfnamefont {J.~M.}\
  \bibnamefont {Tranquada}}, \ and\ \bibinfo {author} {\bibfnamefont {J.~P.}\
  \bibnamefont {Hill}},\ }\bibfield  {title} {\enquote {\bibinfo {title}
  {{X-ray diffuse scattering study of local distortions in Fe$_{1-x}$Te induced
  by excess Fe}},}\ }\href {\doibase 10.1103/PhysRevB.83.184523} {\bibfield
  {journal} {\bibinfo  {journal} {Phys. Rev. B}\ }\textbf {\bibinfo {volume}
  {83}},\ \bibinfo {pages} {184523} (\bibinfo {year} {2011})}\BibitemShut
  {NoStop}%
\bibitem [{\citenamefont {Rodriguez}\ \emph {et~al.}(2011)\citenamefont
  {Rodriguez}, \citenamefont {Stock}, \citenamefont {Zajdel}, \citenamefont
  {Krycka}, \citenamefont {Majkrzak}, \citenamefont {Zavalij},\ and\
  \citenamefont {Green}}]{rodriguez11}%
  \BibitemOpen
  \bibfield  {author} {\bibinfo {author} {\bibfnamefont {E.~E.}\ \bibnamefont
  {Rodriguez}}, \bibinfo {author} {\bibfnamefont {C.}~\bibnamefont {Stock}},
  \bibinfo {author} {\bibfnamefont {P.}~\bibnamefont {Zajdel}}, \bibinfo
  {author} {\bibfnamefont {K.~L.}\ \bibnamefont {Krycka}}, \bibinfo {author}
  {\bibfnamefont {C.~F.}\ \bibnamefont {Majkrzak}}, \bibinfo {author}
  {\bibfnamefont {P.}~\bibnamefont {Zavalij}}, \ and\ \bibinfo {author}
  {\bibfnamefont {M.~A.}\ \bibnamefont {Green}},\ }\bibfield  {title} {\enquote
  {\bibinfo {title} {{Magnetic-crystallographic phase diagram of the
  superconducting parent compound Fe$_{1+x}$Te}},}\ }\href {\doibase
  10.1103/PhysRevB.84.064403} {\bibfield  {journal} {\bibinfo  {journal} {Phys.
  Rev. B}\ }\textbf {\bibinfo {volume} {84}},\ \bibinfo {pages} {064403}
  (\bibinfo {year} {2011})}\BibitemShut {NoStop}%
\bibitem [{\citenamefont {Li}\ \emph {et~al.}(2009)\citenamefont {Li},
  \citenamefont {de~la Cruz}, \citenamefont {Huang}, \citenamefont {Chen},
  \citenamefont {Lynn}, \citenamefont {Hu}, \citenamefont {Huang},
  \citenamefont {Hsu}, \citenamefont {Yeh}, \citenamefont {Wu},\ and\
  \citenamefont {Dai}}]{li09}%
  \BibitemOpen
  \bibfield  {author} {\bibinfo {author} {\bibfnamefont {Shiliang}\
  \bibnamefont {Li}}, \bibinfo {author} {\bibfnamefont {Clarina}\ \bibnamefont
  {de~la Cruz}}, \bibinfo {author} {\bibfnamefont {Q.}~\bibnamefont {Huang}},
  \bibinfo {author} {\bibfnamefont {Y.}~\bibnamefont {Chen}}, \bibinfo {author}
  {\bibfnamefont {J.~W.}\ \bibnamefont {Lynn}}, \bibinfo {author}
  {\bibfnamefont {Jiangping}\ \bibnamefont {Hu}}, \bibinfo {author}
  {\bibfnamefont {Yi-Lin}\ \bibnamefont {Huang}}, \bibinfo {author}
  {\bibfnamefont {Fong-Chi}\ \bibnamefont {Hsu}}, \bibinfo {author}
  {\bibfnamefont {Kuo-Wei}\ \bibnamefont {Yeh}}, \bibinfo {author}
  {\bibfnamefont {Maw-Kuen}\ \bibnamefont {Wu}}, \ and\ \bibinfo {author}
  {\bibfnamefont {Pengcheng}\ \bibnamefont {Dai}},\ }\bibfield  {title}
  {\enquote {\bibinfo {title} {{First-order magnetic and structural phase
  transitions in
  ${\text{Fe}}_{1+y}{\text{Se}}_{x}{\text{Te}}_{1\ensuremath{-}x}$}},}\ }\href
  {\doibase 10.1103/PhysRevB.79.054503} {\bibfield  {journal} {\bibinfo
  {journal} {Phys. Rev. B}\ }\textbf {\bibinfo {volume} {79}},\ \bibinfo
  {pages} {054503} (\bibinfo {year} {2009})}\BibitemShut {NoStop}%
\bibitem [{\citenamefont {Liu}\ \emph {et~al.}(2010)\citenamefont {Liu},
  \citenamefont {Hu}, \citenamefont {Qian}, \citenamefont {Fobes},
  \citenamefont {Mao}, \citenamefont {Bao}, \citenamefont {Reehuis},
  \citenamefont {Kimber}, \citenamefont {Proke\v{s}}, \citenamefont {Matas},
  \citenamefont {Argyriou}, \citenamefont {Hiess}, \citenamefont {Rotaru},
  \citenamefont {Pham}, \citenamefont {Spinu}, \citenamefont {Qiu},
  \citenamefont {Thampy}, \citenamefont {Savici}, \citenamefont {Rodriguez},\
  and\ \citenamefont {Broholm}}]{liu10}%
  \BibitemOpen
  \bibfield  {author} {\bibinfo {author} {\bibfnamefont {T.~J.}\ \bibnamefont
  {Liu}}, \bibinfo {author} {\bibfnamefont {J.}~\bibnamefont {Hu}}, \bibinfo
  {author} {\bibfnamefont {B.}~\bibnamefont {Qian}}, \bibinfo {author}
  {\bibfnamefont {D.}~\bibnamefont {Fobes}}, \bibinfo {author} {\bibfnamefont
  {Z.~Q.}\ \bibnamefont {Mao}}, \bibinfo {author} {\bibfnamefont
  {W.}~\bibnamefont {Bao}}, \bibinfo {author} {\bibfnamefont {M.}~\bibnamefont
  {Reehuis}}, \bibinfo {author} {\bibfnamefont {S.~A.~J.}\ \bibnamefont
  {Kimber}}, \bibinfo {author} {\bibfnamefont {K.}~\bibnamefont {Proke\v{s}}},
  \bibinfo {author} {\bibfnamefont {S.}~\bibnamefont {Matas}}, \bibinfo
  {author} {\bibfnamefont {D.~N.}\ \bibnamefont {Argyriou}}, \bibinfo {author}
  {\bibfnamefont {A.}~\bibnamefont {Hiess}}, \bibinfo {author} {\bibfnamefont
  {A.}~\bibnamefont {Rotaru}}, \bibinfo {author} {\bibfnamefont
  {H.}~\bibnamefont {Pham}}, \bibinfo {author} {\bibfnamefont {L.}~\bibnamefont
  {Spinu}}, \bibinfo {author} {\bibfnamefont {Y.}~\bibnamefont {Qiu}}, \bibinfo
  {author} {\bibfnamefont {V.}~\bibnamefont {Thampy}}, \bibinfo {author}
  {\bibfnamefont {A.~T.}\ \bibnamefont {Savici}}, \bibinfo {author}
  {\bibfnamefont {J.~A.}\ \bibnamefont {Rodriguez}}, \ and\ \bibinfo {author}
  {\bibfnamefont {C.}~\bibnamefont {Broholm}},\ }\bibfield  {title} {\enquote
  {\bibinfo {title} {From $(\pi,0)$ magnetic order to superconductivity with
  $(\pi,\pi)$ magnetic resonance in {Fe}$_{1.02}${Te}$_{1-x}${Se}$_x$},}\
  }\href {\doibase 10.1038/nmat2800} {\bibfield  {journal} {\bibinfo  {journal}
  {Nat. Mater.}\ }\textbf {\bibinfo {volume} {9}},\ \bibinfo {pages} {718--720}
  (\bibinfo {year} {2010})}\BibitemShut {NoStop}%
\bibitem [{\citenamefont {Fang}\ \emph {et~al.}(2008)\citenamefont {Fang},
  \citenamefont {Pham}, \citenamefont {Qian}, \citenamefont {Liu},
  \citenamefont {Vehstedt}, \citenamefont {Liu}, \citenamefont {Spinu},\ and\
  \citenamefont {Mao}}]{fang08}%
  \BibitemOpen
  \bibfield  {author} {\bibinfo {author} {\bibfnamefont {M.~H.}\ \bibnamefont
  {Fang}}, \bibinfo {author} {\bibfnamefont {H.~M.}\ \bibnamefont {Pham}},
  \bibinfo {author} {\bibfnamefont {B.}~\bibnamefont {Qian}}, \bibinfo {author}
  {\bibfnamefont {T.~J.}\ \bibnamefont {Liu}}, \bibinfo {author} {\bibfnamefont
  {E.~K.}\ \bibnamefont {Vehstedt}}, \bibinfo {author} {\bibfnamefont
  {Y.}~\bibnamefont {Liu}}, \bibinfo {author} {\bibfnamefont {L.}~\bibnamefont
  {Spinu}}, \ and\ \bibinfo {author} {\bibfnamefont {Z.~Q.}\ \bibnamefont
  {Mao}},\ }\bibfield  {title} {\enquote {\bibinfo {title} {Superconductivity
  close to magnetic instability in
  $\text{Fe}{({\text{Se}}_{1\ensuremath{-}x}{\text{Te}}_{x})}_{0.82}$},}\
  }\href {\doibase 10.1103/PhysRevB.78.224503} {\bibfield  {journal} {\bibinfo
  {journal} {Phys. Rev. B}\ }\textbf {\bibinfo {volume} {78}},\ \bibinfo
  {pages} {224503} (\bibinfo {year} {2008})}\BibitemShut {NoStop}%
\bibitem [{\citenamefont {Taen}\ \emph {et~al.}(2009)\citenamefont {Taen},
  \citenamefont {Tsuchiya}, \citenamefont {Nakajima},\ and\ \citenamefont
  {Tamegai}}]{taen09}%
  \BibitemOpen
  \bibfield  {author} {\bibinfo {author} {\bibfnamefont {T.}~\bibnamefont
  {Taen}}, \bibinfo {author} {\bibfnamefont {Y.}~\bibnamefont {Tsuchiya}},
  \bibinfo {author} {\bibfnamefont {Y.}~\bibnamefont {Nakajima}}, \ and\
  \bibinfo {author} {\bibfnamefont {T.}~\bibnamefont {Tamegai}},\ }\bibfield
  {title} {\enquote {\bibinfo {title} {{Superconductivity at $T_{c}\sim 14$~K
  in single-crystalline FeTe$_{0.61}$Se$_{0.39}$}},}\ }\href {\doibase
  10.1103/PhysRevB.80.092502} {\bibfield  {journal} {\bibinfo  {journal} {Phys.
  Rev. B}\ }\textbf {\bibinfo {volume} {80}},\ \bibinfo {pages} {092502}
  (\bibinfo {year} {2009})}\BibitemShut {NoStop}%
\bibitem [{\citenamefont {Sales}\ \emph {et~al.}(2009)\citenamefont {Sales},
  \citenamefont {Sefat}, \citenamefont {McGuire}, \citenamefont {Jin},
  \citenamefont {Mandrus},\ and\ \citenamefont {Mozharivskyj}}]{sales09}%
  \BibitemOpen
  \bibfield  {author} {\bibinfo {author} {\bibfnamefont {B.~C.}\ \bibnamefont
  {Sales}}, \bibinfo {author} {\bibfnamefont {A.~S.}\ \bibnamefont {Sefat}},
  \bibinfo {author} {\bibfnamefont {M.~A.}\ \bibnamefont {McGuire}}, \bibinfo
  {author} {\bibfnamefont {R.~Y.}\ \bibnamefont {Jin}}, \bibinfo {author}
  {\bibfnamefont {D.}~\bibnamefont {Mandrus}}, \ and\ \bibinfo {author}
  {\bibfnamefont {Y.}~\bibnamefont {Mozharivskyj}},\ }\bibfield  {title}
  {\enquote {\bibinfo {title} {{Bulk superconductivity at 14 K in single
  crystals of Fe$_{1+y}$Te$_x$Se$_{1-x}$}},}\ }\href {\doibase
  10.1103/PhysRevB.79.094521} {\bibfield  {journal} {\bibinfo  {journal} {Phys.
  Rev. B}\ }\textbf {\bibinfo {volume} {79}},\ \bibinfo {pages} {094521}
  (\bibinfo {year} {2009})}\BibitemShut {NoStop}%
\bibitem [{\citenamefont {Chen}\ \emph {et~al.}(2009)\citenamefont {Chen},
  \citenamefont {Chen}, \citenamefont {Dong}, \citenamefont {Hu}, \citenamefont
  {Li}, \citenamefont {Zhang}, \citenamefont {Zheng}, \citenamefont {Luo},\
  and\ \citenamefont {Wang}}]{chen09}%
  \BibitemOpen
  \bibfield  {author} {\bibinfo {author} {\bibfnamefont {G.~F.}\ \bibnamefont
  {Chen}}, \bibinfo {author} {\bibfnamefont {Z.~G.}\ \bibnamefont {Chen}},
  \bibinfo {author} {\bibfnamefont {J.}~\bibnamefont {Dong}}, \bibinfo {author}
  {\bibfnamefont {W.~Z.}\ \bibnamefont {Hu}}, \bibinfo {author} {\bibfnamefont
  {G.}~\bibnamefont {Li}}, \bibinfo {author} {\bibfnamefont {X.~D.}\
  \bibnamefont {Zhang}}, \bibinfo {author} {\bibfnamefont {P.}~\bibnamefont
  {Zheng}}, \bibinfo {author} {\bibfnamefont {J.~L.}\ \bibnamefont {Luo}}, \
  and\ \bibinfo {author} {\bibfnamefont {N.~L.}\ \bibnamefont {Wang}},\
  }\bibfield  {title} {\enquote {\bibinfo {title} {{Electronic properties of
  single-crystalline Fe$_{1.05}$Te and Fe$_{1.03}$Se$_{0.30}$Te$_{0.70}$}},}\
  }\href {\doibase 10.1103/PhysRevB.79.140509} {\bibfield  {journal} {\bibinfo
  {journal} {Phys. Rev. B}\ }\textbf {\bibinfo {volume} {79}},\ \bibinfo
  {pages} {140509(R)} (\bibinfo {year} {2009})}\BibitemShut {NoStop}%
\bibitem [{\citenamefont {Hancock}\ \emph {et~al.}(2010)\citenamefont
  {Hancock}, \citenamefont {Mirzaei}, \citenamefont {Gillett}, \citenamefont
  {Sebastian}, \citenamefont {Teyssier}, \citenamefont {Viennois},
  \citenamefont {Giannini},\ and\ \citenamefont {van~der Marel}}]{hancock10}%
  \BibitemOpen
  \bibfield  {author} {\bibinfo {author} {\bibfnamefont {J.~N.}\ \bibnamefont
  {Hancock}}, \bibinfo {author} {\bibfnamefont {S.~I.}\ \bibnamefont
  {Mirzaei}}, \bibinfo {author} {\bibfnamefont {J.}~\bibnamefont {Gillett}},
  \bibinfo {author} {\bibfnamefont {S.~E.}\ \bibnamefont {Sebastian}}, \bibinfo
  {author} {\bibfnamefont {J.}~\bibnamefont {Teyssier}}, \bibinfo {author}
  {\bibfnamefont {R.}~\bibnamefont {Viennois}}, \bibinfo {author}
  {\bibfnamefont {E.}~\bibnamefont {Giannini}}, \ and\ \bibinfo {author}
  {\bibfnamefont {D.}~\bibnamefont {van~der Marel}},\ }\bibfield  {title}
  {\enquote {\bibinfo {title} {Strong coupling to magnetic fluctuations in the
  charge dynamics of iron-based superconductors},}\ }\href {\doibase
  10.1103/PhysRevB.82.014523} {\bibfield  {journal} {\bibinfo  {journal} {Phys.
  Rev. B}\ }\textbf {\bibinfo {volume} {82}},\ \bibinfo {pages} {014523}
  (\bibinfo {year} {2010})}\BibitemShut {NoStop}%
\bibitem [{\citenamefont {Dai}\ \emph {et~al.}(2014)\citenamefont {Dai},
  \citenamefont {Akrap}, \citenamefont {Schneeloch}, \citenamefont {Zhong},
  \citenamefont {Liu}, \citenamefont {Gu}, \citenamefont {Li},\ and\
  \citenamefont {Homes}}]{dai14}%
  \BibitemOpen
  \bibfield  {author} {\bibinfo {author} {\bibfnamefont {Y.~M.}\ \bibnamefont
  {Dai}}, \bibinfo {author} {\bibfnamefont {A.}~\bibnamefont {Akrap}}, \bibinfo
  {author} {\bibfnamefont {J.}~\bibnamefont {Schneeloch}}, \bibinfo {author}
  {\bibfnamefont {R.~D.}\ \bibnamefont {Zhong}}, \bibinfo {author}
  {\bibfnamefont {T.~S.}\ \bibnamefont {Liu}}, \bibinfo {author} {\bibfnamefont
  {G.~D.}\ \bibnamefont {Gu}}, \bibinfo {author} {\bibfnamefont
  {Q.}~\bibnamefont {Li}}, \ and\ \bibinfo {author} {\bibfnamefont {C.~C.}\
  \bibnamefont {Homes}},\ }\bibfield  {title} {\enquote {\bibinfo {title}
  {Spectral weight transfer in strongly correlated {Fe}$_{1.03}${Te}},}\ }\href
  {\doibase 10.1103/PhysRevB.90.121114} {\bibfield  {journal} {\bibinfo
  {journal} {Phys. Rev. B}\ }\textbf {\bibinfo {volume} {90}},\ \bibinfo
  {pages} {121114} (\bibinfo {year} {2014})}\BibitemShut {NoStop}%
\bibitem [{\citenamefont {Homes}\ \emph {et~al.}(2010)\citenamefont {Homes},
  \citenamefont {Akrap}, \citenamefont {Wen}, \citenamefont {Xu}, \citenamefont
  {Lin}, \citenamefont {Li},\ and\ \citenamefont {Gu}}]{homes10}%
  \BibitemOpen
  \bibfield  {author} {\bibinfo {author} {\bibfnamefont {C.~C.}\ \bibnamefont
  {Homes}}, \bibinfo {author} {\bibfnamefont {A.}~\bibnamefont {Akrap}},
  \bibinfo {author} {\bibfnamefont {J.~S.}\ \bibnamefont {Wen}}, \bibinfo
  {author} {\bibfnamefont {Z.~J.}\ \bibnamefont {Xu}}, \bibinfo {author}
  {\bibfnamefont {Z.~W.}\ \bibnamefont {Lin}}, \bibinfo {author} {\bibfnamefont
  {Q.}~\bibnamefont {Li}}, \ and\ \bibinfo {author} {\bibfnamefont {G.~D.}\
  \bibnamefont {Gu}},\ }\bibfield  {title} {\enquote {\bibinfo {title}
  {Electronic correlations and unusual superconducting response in the optical
  properties of the iron chalcogenide {FeTe}$_{0.55}${Se}$_{0.45}$},}\ }\href
  {\doibase 10.1103/PhysRevB.81.180508} {\bibfield  {journal} {\bibinfo
  {journal} {Phys. Rev. B}\ }\textbf {\bibinfo {volume} {81}},\ \bibinfo
  {pages} {180508} (\bibinfo {year} {2010})}\BibitemShut {NoStop}%
\bibitem [{\citenamefont {Homes}\ \emph {et~al.}(2015)\citenamefont {Homes},
  \citenamefont {Dai}, \citenamefont {Wen}, \citenamefont {Xu},\ and\
  \citenamefont {Gu}}]{homes15}%
  \BibitemOpen
  \bibfield  {author} {\bibinfo {author} {\bibfnamefont {C.~C.}\ \bibnamefont
  {Homes}}, \bibinfo {author} {\bibfnamefont {Y.~M.}\ \bibnamefont {Dai}},
  \bibinfo {author} {\bibfnamefont {J.~S.}\ \bibnamefont {Wen}}, \bibinfo
  {author} {\bibfnamefont {Z.~J.}\ \bibnamefont {Xu}}, \ and\ \bibinfo {author}
  {\bibfnamefont {G.~D.}\ \bibnamefont {Gu}},\ }\bibfield  {title} {\enquote
  {\bibinfo {title} {{FeTe}$_{0.55}${Se}$_{0.45}$: {A} multiband superconductor
  in the clean and dirty limit},}\ }\href {\doibase 10.1103/PhysRevB.91.144503}
  {\bibfield  {journal} {\bibinfo  {journal} {Phys. Rev. B}\ }\textbf {\bibinfo
  {volume} {91}},\ \bibinfo {pages} {144503} (\bibinfo {year}
  {2015})}\BibitemShut {NoStop}%
\bibitem [{\citenamefont {Pimenov}\ \emph {et~al.}(2013)\citenamefont
  {Pimenov}, \citenamefont {Engelbrecht}, \citenamefont {Shuvaev},
  \citenamefont {Jin}, \citenamefont {Wu}, \citenamefont {Xu}, \citenamefont
  {Cao},\ and\ \citenamefont {Schachinger}}]{pimenov13}%
  \BibitemOpen
  \bibfield  {author} {\bibinfo {author} {\bibfnamefont {A.}~\bibnamefont
  {Pimenov}}, \bibinfo {author} {\bibfnamefont {S.}~\bibnamefont
  {Engelbrecht}}, \bibinfo {author} {\bibfnamefont {A.~M.}\ \bibnamefont
  {Shuvaev}}, \bibinfo {author} {\bibfnamefont {B.~B.}\ \bibnamefont {Jin}},
  \bibinfo {author} {\bibfnamefont {P.~H.}\ \bibnamefont {Wu}}, \bibinfo
  {author} {\bibfnamefont {B.}~\bibnamefont {Xu}}, \bibinfo {author}
  {\bibfnamefont {L.~X.}\ \bibnamefont {Cao}}, \ and\ \bibinfo {author}
  {\bibfnamefont {E.}~\bibnamefont {Schachinger}},\ }\bibfield  {title}
  {\enquote {\bibinfo {title} {{Terahertz conductivity in
  {FeSe}$_{0.5}${Te}$_{0.5}$ superconducting films}},}\ }\href {\doibase
  10.1088/1367-2630/15/1/013032} {\bibfield  {journal} {\bibinfo  {journal}
  {New J. Phys.}\ }\textbf {\bibinfo {volume} {15}},\ \bibinfo {pages} {013032}
  (\bibinfo {year} {2013})}\BibitemShut {NoStop}%
\bibitem [{\citenamefont {Perucchi}\ \emph {et~al.}(2014)\citenamefont
  {Perucchi}, \citenamefont {Joseph}, \citenamefont {Caramazza}, \citenamefont
  {Autore}, \citenamefont {Bellingeri}, \citenamefont {Kawale}, \citenamefont
  {Ferdeghini}, \citenamefont {Putti}, \citenamefont {Lupi},\ and\
  \citenamefont {Dore}}]{perucchi14}%
  \BibitemOpen
  \bibfield  {author} {\bibinfo {author} {\bibfnamefont {A.}~\bibnamefont
  {Perucchi}}, \bibinfo {author} {\bibfnamefont {B.}~\bibnamefont {Joseph}},
  \bibinfo {author} {\bibfnamefont {S.}~\bibnamefont {Caramazza}}, \bibinfo
  {author} {\bibfnamefont {M.}~\bibnamefont {Autore}}, \bibinfo {author}
  {\bibfnamefont {E.}~\bibnamefont {Bellingeri}}, \bibinfo {author}
  {\bibfnamefont {S.}~\bibnamefont {Kawale}}, \bibinfo {author} {\bibfnamefont
  {C.}~\bibnamefont {Ferdeghini}}, \bibinfo {author} {\bibfnamefont
  {M.}~\bibnamefont {Putti}}, \bibinfo {author} {\bibfnamefont
  {S.}~\bibnamefont {Lupi}}, \ and\ \bibinfo {author} {\bibfnamefont
  {P.}~\bibnamefont {Dore}},\ }\bibfield  {title} {\enquote {\bibinfo {title}
  {{Two-band conductivity of a {FeSe}$_{0.5}${Te}$_{0.5}$ film by reflectance
  measurements in the terahertz and infrared range}},}\ }\href {\doibase
  10.1088/0953-2048/27/12/125011} {\bibfield  {journal} {\bibinfo  {journal}
  {Supercond. Sci. Technol.}\ }\textbf {\bibinfo {volume} {27}},\ \bibinfo
  {pages} {125011} (\bibinfo {year} {2014})}\BibitemShut {NoStop}%
\bibitem [{\citenamefont {Xia}\ \emph {et~al.}(2009)\citenamefont {Xia},
  \citenamefont {Hou}, \citenamefont {Zhao}, \citenamefont {Zhang},
  \citenamefont {Chen}, \citenamefont {Luo}, \citenamefont {Wang},
  \citenamefont {Wei}, \citenamefont {Lu},\ and\ \citenamefont
  {Zhang}}]{xia09}%
  \BibitemOpen
  \bibfield  {author} {\bibinfo {author} {\bibfnamefont {T.-L.}\ \bibnamefont
  {Xia}}, \bibinfo {author} {\bibfnamefont {D.}~\bibnamefont {Hou}}, \bibinfo
  {author} {\bibfnamefont {S.~C.}\ \bibnamefont {Zhao}}, \bibinfo {author}
  {\bibfnamefont {A.~M.}\ \bibnamefont {Zhang}}, \bibinfo {author}
  {\bibfnamefont {G.~F.}\ \bibnamefont {Chen}}, \bibinfo {author}
  {\bibfnamefont {J.~L.}\ \bibnamefont {Luo}}, \bibinfo {author} {\bibfnamefont
  {N.~L.}\ \bibnamefont {Wang}}, \bibinfo {author} {\bibfnamefont {J.~H.}\
  \bibnamefont {Wei}}, \bibinfo {author} {\bibfnamefont {Z.-Y.}\ \bibnamefont
  {Lu}}, \ and\ \bibinfo {author} {\bibfnamefont {Q.~M.}\ \bibnamefont
  {Zhang}},\ }\bibfield  {title} {\enquote {\bibinfo {title} {{Raman phonons of
  $\ensuremath{\alpha}\text{-FeTe}$ and
  ${\text{Fe}}_{1.03}{\text{Se}}_{0.3}{\text{Te}}_{0.7}$ single crystals}},}\
  }\href {\doibase 10.1103/PhysRevB.79.140510} {\bibfield  {journal} {\bibinfo
  {journal} {Phys. Rev. B}\ }\textbf {\bibinfo {volume} {79}},\ \bibinfo
  {pages} {140510(R)} (\bibinfo {year} {2009})}\BibitemShut {NoStop}%
\bibitem [{\citenamefont {Okazaki}\ \emph {et~al.}(2011)\citenamefont
  {Okazaki}, \citenamefont {Sugai}, \citenamefont {Niitaka},\ and\
  \citenamefont {Takagi}}]{okazaki11}%
  \BibitemOpen
  \bibfield  {author} {\bibinfo {author} {\bibfnamefont {K.}~\bibnamefont
  {Okazaki}}, \bibinfo {author} {\bibfnamefont {S.}~\bibnamefont {Sugai}},
  \bibinfo {author} {\bibfnamefont {S.}~\bibnamefont {Niitaka}}, \ and\
  \bibinfo {author} {\bibfnamefont {H.}~\bibnamefont {Takagi}},\ }\bibfield
  {title} {\enquote {\bibinfo {title} {{Phonon, two-magnon, and electronic
  Raman scattering of Fe$_{1+y}$Te$_{1-x}$Se$_{x}$}},}\ }\href {\doibase
  10.1103/PhysRevB.83.035103} {\bibfield  {journal} {\bibinfo  {journal} {Phys.
  Rev. B}\ }\textbf {\bibinfo {volume} {83}},\ \bibinfo {pages} {035103}
  (\bibinfo {year} {2011})}\BibitemShut {NoStop}%
\bibitem [{\citenamefont {Gnezdilov}\ \emph {et~al.}(2011)\citenamefont
  {Gnezdilov}, \citenamefont {Pashkevich}, \citenamefont {Lemmens},
  \citenamefont {Gusev}, \citenamefont {Lamonova}, \citenamefont {Shevtsova},
  \citenamefont {Vitebskiy}, \citenamefont {Afanasiev}, \citenamefont
  {Gnatchenko}, \citenamefont {Tsurkan}, \citenamefont {Deisenhofer},\ and\
  \citenamefont {Loidl}}]{gnezdilov11}%
  \BibitemOpen
  \bibfield  {author} {\bibinfo {author} {\bibfnamefont {V.}~\bibnamefont
  {Gnezdilov}}, \bibinfo {author} {\bibfnamefont {Yu.}\ \bibnamefont
  {Pashkevich}}, \bibinfo {author} {\bibfnamefont {P.}~\bibnamefont {Lemmens}},
  \bibinfo {author} {\bibfnamefont {A.}~\bibnamefont {Gusev}}, \bibinfo
  {author} {\bibfnamefont {K.}~\bibnamefont {Lamonova}}, \bibinfo {author}
  {\bibfnamefont {T.}~\bibnamefont {Shevtsova}}, \bibinfo {author}
  {\bibfnamefont {I.}~\bibnamefont {Vitebskiy}}, \bibinfo {author}
  {\bibfnamefont {O.}~\bibnamefont {Afanasiev}}, \bibinfo {author}
  {\bibfnamefont {S.}~\bibnamefont {Gnatchenko}}, \bibinfo {author}
  {\bibfnamefont {V.}~\bibnamefont {Tsurkan}}, \bibinfo {author} {\bibfnamefont
  {J.}~\bibnamefont {Deisenhofer}}, \ and\ \bibinfo {author} {\bibfnamefont
  {A.}~\bibnamefont {Loidl}},\ }\bibfield  {title} {\enquote {\bibinfo {title}
  {{Anomalous optical phonons in FeTe chalcogenides: Spin state, magnetic
  order, and lattice anharmonicity}},}\ }\href {\doibase
  10.1103/PhysRevB.83.245127} {\bibfield  {journal} {\bibinfo  {journal} {Phys.
  Rev. B}\ }\textbf {\bibinfo {volume} {83}},\ \bibinfo {pages} {245127}
  (\bibinfo {year} {2011})}\BibitemShut {NoStop}%
\bibitem [{\citenamefont {Um}\ \emph {et~al.}(2012)\citenamefont {Um},
  \citenamefont {Subedi}, \citenamefont {Toulemonde}, \citenamefont {Ganin},
  \citenamefont {Boeri}, \citenamefont {Rahlenbeck}, \citenamefont {Liu},
  \citenamefont {Lin}, \citenamefont {Carlsson}, \citenamefont {Sulpice},
  \citenamefont {Rosseinsky}, \citenamefont {Keimer},\ and\ \citenamefont
  {Le~Tacon}}]{um12}%
  \BibitemOpen
  \bibfield  {author} {\bibinfo {author} {\bibfnamefont {Y.~J.}\ \bibnamefont
  {Um}}, \bibinfo {author} {\bibfnamefont {A.}~\bibnamefont {Subedi}}, \bibinfo
  {author} {\bibfnamefont {P.}~\bibnamefont {Toulemonde}}, \bibinfo {author}
  {\bibfnamefont {A.~Y.}\ \bibnamefont {Ganin}}, \bibinfo {author}
  {\bibfnamefont {L.}~\bibnamefont {Boeri}}, \bibinfo {author} {\bibfnamefont
  {M.}~\bibnamefont {Rahlenbeck}}, \bibinfo {author} {\bibfnamefont
  {Y.}~\bibnamefont {Liu}}, \bibinfo {author} {\bibfnamefont {C.~T.}\
  \bibnamefont {Lin}}, \bibinfo {author} {\bibfnamefont {S.~J.~E.}\
  \bibnamefont {Carlsson}}, \bibinfo {author} {\bibfnamefont {A.}~\bibnamefont
  {Sulpice}}, \bibinfo {author} {\bibfnamefont {M.~J.}\ \bibnamefont
  {Rosseinsky}}, \bibinfo {author} {\bibfnamefont {B.}~\bibnamefont {Keimer}},
  \ and\ \bibinfo {author} {\bibfnamefont {M.}~\bibnamefont {Le~Tacon}},\
  }\bibfield  {title} {\enquote {\bibinfo {title} {{Anomalous dependence of
  $c$-axis polarized Fe ${\mathrm{B}}_{1g}$ phonon mode with Fe and Se
  concentrations in Fe$_{1+y}$Te$_{1\ensuremath{-}x}$Se$_{x}$}},}\ }\href
  {\doibase 10.1103/PhysRevB.85.064519} {\bibfield  {journal} {\bibinfo
  {journal} {Phys. Rev. B}\ }\textbf {\bibinfo {volume} {85}},\ \bibinfo
  {pages} {064519} (\bibinfo {year} {2012})}\BibitemShut {NoStop}%
\bibitem [{\citenamefont {Popovi\'{c}}\ \emph {et~al.}(2014)\citenamefont
  {Popovi\'{c}}, \citenamefont {Lazarevi\'{c}}, \citenamefont {Bogdanovi\'{c}},
  \citenamefont {Radonji\'{c}}, \citenamefont {Tanaskovi\'{c}}, \citenamefont
  {Hu}, \citenamefont {Lei},\ and\ \citenamefont {Petrovic}}]{popovic14}%
  \BibitemOpen
  \bibfield  {author} {\bibinfo {author} {\bibfnamefont {Z.~V.}\ \bibnamefont
  {Popovi\'{c}}}, \bibinfo {author} {\bibfnamefont {N.}~\bibnamefont
  {Lazarevi\'{c}}}, \bibinfo {author} {\bibfnamefont {S.}~\bibnamefont
  {Bogdanovi\'{c}}}, \bibinfo {author} {\bibfnamefont {M.~M.}\ \bibnamefont
  {Radonji\'{c}}}, \bibinfo {author} {\bibfnamefont {D.}~\bibnamefont
  {Tanaskovi\'{c}}}, \bibinfo {author} {\bibfnamefont {Rongwei}\ \bibnamefont
  {Hu}}, \bibinfo {author} {\bibfnamefont {Hechang}\ \bibnamefont {Lei}}, \
  and\ \bibinfo {author} {\bibfnamefont {C.}~\bibnamefont {Petrovic}},\
  }\bibfield  {title} {\enquote {\bibinfo {title} {{Signatures of the
  spin-phonon coupling in alloys}},}\ }\href {\doibase
  10.1016/j.ssc.2014.05.025} {\bibfield  {journal} {\bibinfo  {journal} {Solid
  State Commun.}\ }\textbf {\bibinfo {volume} {193}},\ \bibinfo {pages}
  {51--55} (\bibinfo {year} {2014})}\BibitemShut {NoStop}%
\bibitem [{\citenamefont {Homes}\ \emph {et~al.}(1993)\citenamefont {Homes},
  \citenamefont {Reedyk}, \citenamefont {Cradles},\ and\ \citenamefont
  {Timusk}}]{homes93}%
  \BibitemOpen
  \bibfield  {author} {\bibinfo {author} {\bibfnamefont {C.~C.}\ \bibnamefont
  {Homes}}, \bibinfo {author} {\bibfnamefont {M.}~\bibnamefont {Reedyk}},
  \bibinfo {author} {\bibfnamefont {D.~A.}\ \bibnamefont {Cradles}}, \ and\
  \bibinfo {author} {\bibfnamefont {T.}~\bibnamefont {Timusk}},\ }\bibfield
  {title} {\enquote {\bibinfo {title} {Technique for measuring the reflectance
  of irregular, submillimeter-sized samples},}\ }\href {\doibase
  10.1364/AO.32.002976} {\bibfield  {journal} {\bibinfo  {journal} {Appl.
  Opt.}\ }\textbf {\bibinfo {volume} {32}},\ \bibinfo {pages} {2976--2983}
  (\bibinfo {year} {1993})}\BibitemShut {NoStop}%
\bibitem [{\citenamefont {Dressel}\ and\ \citenamefont
  {Gr{\"u}ner}(2001)}]{dressel-book}%
  \BibitemOpen
  \bibfield  {author} {\bibinfo {author} {\bibfnamefont {M.}~\bibnamefont
  {Dressel}}\ and\ \bibinfo {author} {\bibfnamefont {G.}~\bibnamefont
  {Gr{\"u}ner}},\ }\href@noop {} {\emph {\bibinfo {title} {Electrodynamics of
  Solids}}}\ (\bibinfo  {publisher} {Cambridge University Press},\ \bibinfo
  {address} {Cambridge},\ \bibinfo {year} {2001})\BibitemShut {NoStop}%
\bibitem [{\citenamefont {Fano}(1961)}]{fano61}%
  \BibitemOpen
  \bibfield  {author} {\bibinfo {author} {\bibfnamefont {U.}~\bibnamefont
  {Fano}},\ }\bibfield  {title} {\enquote {\bibinfo {title} {Effects of
  configuration interaction on intensities and phase shifts},}\ }\href
  {\doibase 10.1103/PhysRev.124.1866} {\bibfield  {journal} {\bibinfo
  {journal} {Phys. Rev.}\ }\textbf {\bibinfo {volume} {124}},\ \bibinfo {pages}
  {1866--1878} (\bibinfo {year} {1961})}\BibitemShut {NoStop}%
\bibitem [{\citenamefont {Damascelli}(1996)}]{damascelli96}%
  \BibitemOpen
  \bibfield  {author} {\bibinfo {author} {\bibfnamefont {A.}~\bibnamefont
  {Damascelli}},\ }\emph {\bibinfo {title} {Optical Spectroscopy of Quantum
  Spin Systems}},\ \href@noop {} {Ph.D. thesis},\ \bibinfo  {school}
  {University of Groningen} (\bibinfo {year} {1996}),\ \bibinfo {note}
  {p.~21}\BibitemShut {NoStop}%
\bibitem [{ref()}]{reffit}%
  \BibitemOpen
  \href@noop {} {}\bibinfo {note} {A. Kuzmenko, Software RefFIT, Manual p.~64
  (2014)}\BibitemShut {NoStop}%
\bibitem [{\citenamefont {Subedi}\ \emph {et~al.}(2008)\citenamefont {Subedi},
  \citenamefont {Zhang}, \citenamefont {Singh},\ and\ \citenamefont
  {Du}}]{subedi08}%
  \BibitemOpen
  \bibfield  {author} {\bibinfo {author} {\bibfnamefont {Alaska}\ \bibnamefont
  {Subedi}}, \bibinfo {author} {\bibfnamefont {Lijun}\ \bibnamefont {Zhang}},
  \bibinfo {author} {\bibfnamefont {D.~J.}\ \bibnamefont {Singh}}, \ and\
  \bibinfo {author} {\bibfnamefont {M.~H.}\ \bibnamefont {Du}},\ }\bibfield
  {title} {\enquote {\bibinfo {title} {Density functional study of {FeS},
  {FeSe}, and {FeTe}: Electronic structure, magnetism, phonons, and
  superconductivity},}\ }\href {\doibase 10.1103/PhysRevB.78.134514} {\bibfield
   {journal} {\bibinfo  {journal} {Phys. Rev. B}\ }\textbf {\bibinfo {volume}
  {78}},\ \bibinfo {pages} {134514} (\bibinfo {year} {2008})}\BibitemShut
  {NoStop}%
\bibitem [{\citenamefont {Wu}\ \emph {et~al.}(2010)\citenamefont {Wu},
  \citenamefont {Bari{\v{s}}i{\'{c}}}, \citenamefont {Kallina}, \citenamefont
  {Faridian}, \citenamefont {Gorshunov}, \citenamefont {Drichko}, \citenamefont
  {Li}, \citenamefont {Lin}, \citenamefont {Cao}, \citenamefont {Xu},
  \citenamefont {Wang},\ and\ \citenamefont {Dressel}}]{wu10a}%
  \BibitemOpen
  \bibfield  {author} {\bibinfo {author} {\bibfnamefont {D.}~\bibnamefont
  {Wu}}, \bibinfo {author} {\bibfnamefont {N.}~\bibnamefont
  {Bari{\v{s}}i{\'{c}}}}, \bibinfo {author} {\bibfnamefont {P.}~\bibnamefont
  {Kallina}}, \bibinfo {author} {\bibfnamefont {A.}~\bibnamefont {Faridian}},
  \bibinfo {author} {\bibfnamefont {B.}~\bibnamefont {Gorshunov}}, \bibinfo
  {author} {\bibfnamefont {N.}~\bibnamefont {Drichko}}, \bibinfo {author}
  {\bibfnamefont {L.~J.}\ \bibnamefont {Li}}, \bibinfo {author} {\bibfnamefont
  {X.}~\bibnamefont {Lin}}, \bibinfo {author} {\bibfnamefont {G.~H.}\
  \bibnamefont {Cao}}, \bibinfo {author} {\bibfnamefont {Z.~A.}\ \bibnamefont
  {Xu}}, \bibinfo {author} {\bibfnamefont {N.~L.}\ \bibnamefont {Wang}}, \ and\
  \bibinfo {author} {\bibfnamefont {M.}~\bibnamefont {Dressel}},\ }\bibfield
  {title} {\enquote {\bibinfo {title} {Optical investigations of the normal and
  superconducting states reveal two electronic subsystems in iron pnictides},}\
  }\href {\doibase 10.1103/PhysRevB.81.100512} {\bibfield  {journal} {\bibinfo
  {journal} {Phys. Rev. B}\ }\textbf {\bibinfo {volume} {81}},\ \bibinfo
  {pages} {100512(R)} (\bibinfo {year} {2010})}\BibitemShut {NoStop}%
\bibitem [{\citenamefont {Zhang}\ \emph {et~al.}(2010)\citenamefont {Zhang},
  \citenamefont {Chen}, \citenamefont {He}, \citenamefont {Yang}, \citenamefont
  {Xie}, \citenamefont {Xie}, \citenamefont {Chen}, \citenamefont {Fang},
  \citenamefont {Arita}, \citenamefont {Shimada}, \citenamefont {Namatame},
  \citenamefont {Taniguchi}, \citenamefont {Hu},\ and\ \citenamefont
  {Feng}}]{zhang10}%
  \BibitemOpen
  \bibfield  {author} {\bibinfo {author} {\bibfnamefont {Y.}~\bibnamefont
  {Zhang}}, \bibinfo {author} {\bibfnamefont {F.}~\bibnamefont {Chen}},
  \bibinfo {author} {\bibfnamefont {C.}~\bibnamefont {He}}, \bibinfo {author}
  {\bibfnamefont {L.~X.}\ \bibnamefont {Yang}}, \bibinfo {author}
  {\bibfnamefont {B.~P.}\ \bibnamefont {Xie}}, \bibinfo {author} {\bibfnamefont
  {Y.~L.}\ \bibnamefont {Xie}}, \bibinfo {author} {\bibfnamefont {X.~H.}\
  \bibnamefont {Chen}}, \bibinfo {author} {\bibfnamefont {Minghu}\ \bibnamefont
  {Fang}}, \bibinfo {author} {\bibfnamefont {M.}~\bibnamefont {Arita}},
  \bibinfo {author} {\bibfnamefont {K.}~\bibnamefont {Shimada}}, \bibinfo
  {author} {\bibfnamefont {H.}~\bibnamefont {Namatame}}, \bibinfo {author}
  {\bibfnamefont {M.}~\bibnamefont {Taniguchi}}, \bibinfo {author}
  {\bibfnamefont {J.~P.}\ \bibnamefont {Hu}}, \ and\ \bibinfo {author}
  {\bibfnamefont {D.~L.}\ \bibnamefont {Feng}},\ }\bibfield  {title} {\enquote
  {\bibinfo {title} {{Strong correlations and spin-density-wave phase induced
  by a massive spectral weight redistribution in $\alpha${-Fe}$_{1.06}${Te}
  }},}\ }\href {\doibase 10.1103/PhysRevB.82.165113} {\bibfield  {journal}
  {\bibinfo  {journal} {Phys. Rev. B}\ }\textbf {\bibinfo {volume} {82}},\
  \bibinfo {pages} {165113} (\bibinfo {year} {2010})}\BibitemShut {NoStop}%
\bibitem [{\citenamefont {Lin}\ \emph {et~al.}(2013)\citenamefont {Lin},
  \citenamefont {Texier}, \citenamefont {Taleb-Ibrahimi}, \citenamefont
  {Le~F\`evre}, \citenamefont {Bertran}, \citenamefont {Giannini},
  \citenamefont {Grioni},\ and\ \citenamefont {Brouet}}]{lin13}%
  \BibitemOpen
  \bibfield  {author} {\bibinfo {author} {\bibfnamefont {Ping-Hui}\
  \bibnamefont {Lin}}, \bibinfo {author} {\bibfnamefont {Y.}~\bibnamefont
  {Texier}}, \bibinfo {author} {\bibfnamefont {A.}~\bibnamefont
  {Taleb-Ibrahimi}}, \bibinfo {author} {\bibfnamefont {P.}~\bibnamefont
  {Le~F\`evre}}, \bibinfo {author} {\bibfnamefont {F.}~\bibnamefont {Bertran}},
  \bibinfo {author} {\bibfnamefont {E.}~\bibnamefont {Giannini}}, \bibinfo
  {author} {\bibfnamefont {M.}~\bibnamefont {Grioni}}, \ and\ \bibinfo {author}
  {\bibfnamefont {V.}~\bibnamefont {Brouet}},\ }\bibfield  {title} {\enquote
  {\bibinfo {title} {Nature of the bad metallic behavior of {Fe}$_{1.06}${Te}
  inferred from its evolution in the magnetic state},}\ }\href {\doibase
  10.1103/PhysRevLett.111.217002} {\bibfield  {journal} {\bibinfo  {journal}
  {Phys. Rev. Lett.}\ }\textbf {\bibinfo {volume} {111}},\ \bibinfo {pages}
  {217002} (\bibinfo {year} {2013})}\BibitemShut {NoStop}%
\bibitem [{\citenamefont {Klemens}(1966)}]{klemens66}%
  \BibitemOpen
  \bibfield  {author} {\bibinfo {author} {\bibfnamefont {P.~G.}\ \bibnamefont
  {Klemens}},\ }\bibfield  {title} {\enquote {\bibinfo {title} {Anharmonic
  decay of optical phonons},}\ }\href {\doibase 10.1103/PhysRev.148.845}
  {\bibfield  {journal} {\bibinfo  {journal} {Phys. Rev.}\ }\textbf {\bibinfo
  {volume} {148}},\ \bibinfo {pages} {845--848} (\bibinfo {year}
  {1966})}\BibitemShut {NoStop}%
\bibitem [{\citenamefont {Men\'endez}\ and\ \citenamefont
  {Cardona}(1984)}]{menendez84}%
  \BibitemOpen
  \bibfield  {author} {\bibinfo {author} {\bibfnamefont {Jos\'e}\ \bibnamefont
  {Men\'endez}}\ and\ \bibinfo {author} {\bibfnamefont {Manuel}\ \bibnamefont
  {Cardona}},\ }\bibfield  {title} {\enquote {\bibinfo {title} {{Temperature
  dependence of the first-order Raman scattering by phonons in Si, Ge, and
  $\ensuremath{\alpha}-\mathrm{S}\mathrm{n}$: Anharmonic effects}},}\ }\href
  {\doibase 10.1103/PhysRevB.29.2051} {\bibfield  {journal} {\bibinfo
  {journal} {Phys. Rev. B}\ }\textbf {\bibinfo {volume} {29}},\ \bibinfo
  {pages} {2051--2059} (\bibinfo {year} {1984})}\BibitemShut {NoStop}%
\bibitem [{\citenamefont {Akrap}\ \emph {et~al.}(2009)\citenamefont {Akrap},
  \citenamefont {Tu}, \citenamefont {Li}, \citenamefont {Cao}, \citenamefont
  {Xu},\ and\ \citenamefont {Homes}}]{akrap09}%
  \BibitemOpen
  \bibfield  {author} {\bibinfo {author} {\bibfnamefont {A.}~\bibnamefont
  {Akrap}}, \bibinfo {author} {\bibfnamefont {J.~J.}\ \bibnamefont {Tu}},
  \bibinfo {author} {\bibfnamefont {L.~J.}\ \bibnamefont {Li}}, \bibinfo
  {author} {\bibfnamefont {G.~H.}\ \bibnamefont {Cao}}, \bibinfo {author}
  {\bibfnamefont {Z.~A.}\ \bibnamefont {Xu}}, \ and\ \bibinfo {author}
  {\bibfnamefont {C.~C.}\ \bibnamefont {Homes}},\ }\bibfield  {title} {\enquote
  {\bibinfo {title} {{Infrared phonon anomaly in
  ${\text{BaFe}}_{2}{\text{As}}_{2}$}},}\ }\href {\doibase
  10.1103/PhysRevB.80.180502} {\bibfield  {journal} {\bibinfo  {journal} {Phys.
  Rev. B}\ }\textbf {\bibinfo {volume} {80}},\ \bibinfo {pages} {180502}
  (\bibinfo {year} {2009})}\BibitemShut {NoStop}%
\bibitem [{vib()}]{vibratz}%
  \BibitemOpen
  \href@noop {} {}\bibinfo {note} {E. Dowty, Software VIBRATZ (Shape Software,
  Kingsport, TN, 2001)}\BibitemShut {NoStop}%
\bibitem [{\citenamefont {Tsukada}\ \emph {et~al.}(2010)\citenamefont
  {Tsukada}, \citenamefont {Hanawa}, \citenamefont {Komiya}, \citenamefont
  {Akiike}, \citenamefont {Tanaka}, \citenamefont {Imai},\ and\ \citenamefont
  {Maeda}}]{tsukada10}%
  \BibitemOpen
  \bibfield  {author} {\bibinfo {author} {\bibfnamefont {I.}~\bibnamefont
  {Tsukada}}, \bibinfo {author} {\bibfnamefont {M.}~\bibnamefont {Hanawa}},
  \bibinfo {author} {\bibfnamefont {Seiki}\ \bibnamefont {Komiya}}, \bibinfo
  {author} {\bibfnamefont {T.}~\bibnamefont {Akiike}}, \bibinfo {author}
  {\bibfnamefont {R.}~\bibnamefont {Tanaka}}, \bibinfo {author} {\bibfnamefont
  {Y.}~\bibnamefont {Imai}}, \ and\ \bibinfo {author} {\bibfnamefont
  {A.}~\bibnamefont {Maeda}},\ }\bibfield  {title} {\enquote {\bibinfo {title}
  {{Hall effect in superconducting
  $\text{Fe}({\text{Se}}_{0.5}{\text{Te}}_{0.5})$ thin films}},}\ }\href
  {\doibase 10.1103/PhysRevB.81.054515} {\bibfield  {journal} {\bibinfo
  {journal} {Phys. Rev. B}\ }\textbf {\bibinfo {volume} {81}},\ \bibinfo
  {pages} {054515} (\bibinfo {year} {2010})}\BibitemShut {NoStop}%
\bibitem [{\citenamefont {Luo}\ \emph {et~al.}(2012)\citenamefont {Luo},
  \citenamefont {Wu}, \citenamefont {Cheng}, \citenamefont {Lin}, \citenamefont
  {Wu}, \citenamefont {Uen}, \citenamefont {Juang}, \citenamefont {Kobayashi},
  \citenamefont {Wen}, \citenamefont {Huang}, \citenamefont {Yeh},
  \citenamefont {Wu}, \citenamefont {Chareev}, \citenamefont {Volkova},\ and\
  \citenamefont {Vasiliev}}]{luo12}%
  \BibitemOpen
  \bibfield  {author} {\bibinfo {author} {\bibfnamefont {C.~W.}\ \bibnamefont
  {Luo}}, \bibinfo {author} {\bibfnamefont {I.~H.}\ \bibnamefont {Wu}},
  \bibinfo {author} {\bibfnamefont {P.~C.}\ \bibnamefont {Cheng}}, \bibinfo
  {author} {\bibfnamefont {J.-Y.}\ \bibnamefont {Lin}}, \bibinfo {author}
  {\bibfnamefont {K.~H.}\ \bibnamefont {Wu}}, \bibinfo {author} {\bibfnamefont
  {T.~M.}\ \bibnamefont {Uen}}, \bibinfo {author} {\bibfnamefont {J.~Y.}\
  \bibnamefont {Juang}}, \bibinfo {author} {\bibfnamefont {T.}~\bibnamefont
  {Kobayashi}}, \bibinfo {author} {\bibfnamefont {Y.~C.}\ \bibnamefont {Wen}},
  \bibinfo {author} {\bibfnamefont {T.~W}\ \bibnamefont {Huang}}, \bibinfo
  {author} {\bibfnamefont {K.~W.}\ \bibnamefont {Yeh}}, \bibinfo {author}
  {\bibfnamefont {M.~K.}\ \bibnamefont {Wu}}, \bibinfo {author} {\bibfnamefont
  {D.~A.}\ \bibnamefont {Chareev}}, \bibinfo {author} {\bibfnamefont {O.~S.}\
  \bibnamefont {Volkova}}, \ and\ \bibinfo {author} {\bibfnamefont {A.~N.}\
  \bibnamefont {Vasiliev}},\ }\bibfield  {title} {\enquote {\bibinfo {title}
  {{Ultrafast dynamics and phonon softening in Fe$_{1+y}$Se$_{1-x}$Te$_x$
  single crystals}},}\ }\href {\doibase 0.1088/1367-2630/14/10/103053}
  {\bibfield  {journal} {\bibinfo  {journal} {New J. Phys.}\ }\textbf {\bibinfo
  {volume} {14}},\ \bibinfo {pages} {103053} (\bibinfo {year}
  {2012})}\BibitemShut {NoStop}%
\bibitem [{\citenamefont {Boeri}\ \emph {et~al.}(2008)\citenamefont {Boeri},
  \citenamefont {Dolgov},\ and\ \citenamefont {Golubov}}]{boeri08}%
  \BibitemOpen
  \bibfield  {author} {\bibinfo {author} {\bibfnamefont {L.}~\bibnamefont
  {Boeri}}, \bibinfo {author} {\bibfnamefont {O.~V.}\ \bibnamefont {Dolgov}}, \
  and\ \bibinfo {author} {\bibfnamefont {A.~A.}\ \bibnamefont {Golubov}},\
  }\bibfield  {title} {\enquote {\bibinfo {title} {{Is LaFeAsO$_{1-x}$F$_x$ an
  Electron-Phonon Superconductor?}}}\ }\href {\doibase
  10.1103/PhysRevLett.101.026403} {\bibfield  {journal} {\bibinfo  {journal}
  {Phys. Rev. Lett.}\ }\textbf {\bibinfo {volume} {101}},\ \bibinfo {pages}
  {026403} (\bibinfo {year} {2008})}\BibitemShut {NoStop}%
\bibitem [{\citenamefont {Kontani}\ and\ \citenamefont
  {Onari}(2010)}]{kotani10}%
  \BibitemOpen
  \bibfield  {author} {\bibinfo {author} {\bibfnamefont {Hiroshi}\ \bibnamefont
  {Kontani}}\ and\ \bibinfo {author} {\bibfnamefont {Seiichiro}\ \bibnamefont
  {Onari}},\ }\bibfield  {title} {\enquote {\bibinfo {title}
  {{Orbital-Fluctuation-Mediated Superconductivity in Iron Pnictides: Analysis
  of the Five-Orbital Hubbard-Holstein Model}},}\ }\href {\doibase
  10.1103/PhysRevLett.104.157001} {\bibfield  {journal} {\bibinfo  {journal}
  {Phys. Rev. Lett.}\ }\textbf {\bibinfo {volume} {104}},\ \bibinfo {pages}
  {157001} (\bibinfo {year} {2010})}\BibitemShut {NoStop}%
\bibitem [{\citenamefont {Saito}\ \emph {et~al.}(2010)\citenamefont {Saito},
  \citenamefont {Onari},\ and\ \citenamefont {Kontani}}]{saito10}%
  \BibitemOpen
  \bibfield  {author} {\bibinfo {author} {\bibfnamefont {Tetsuro}\ \bibnamefont
  {Saito}}, \bibinfo {author} {\bibfnamefont {Seiichiro}\ \bibnamefont
  {Onari}}, \ and\ \bibinfo {author} {\bibfnamefont {Hiroshi}\ \bibnamefont
  {Kontani}},\ }\bibfield  {title} {\enquote {\bibinfo {title} {Orbital
  fluctuation theory in iron pnictides: {E}ffects of {As-Fe-As} bond angle,
  isotope substitution, and ${Z}^{2}$-orbital pocket on superconductivity},}\
  }\href {\doibase 10.1103/PhysRevB.82.144510} {\bibfield  {journal} {\bibinfo
  {journal} {Phys. Rev. B}\ }\textbf {\bibinfo {volume} {82}},\ \bibinfo
  {pages} {144510} (\bibinfo {year} {2010})}\BibitemShut {NoStop}%
\bibitem [{\citenamefont {Ma}\ \emph {et~al.}(2014)\citenamefont {Ma},
  \citenamefont {Wu}, \citenamefont {Yin}, \citenamefont {Yang}, \citenamefont
  {Shi}, \citenamefont {Wang}, \citenamefont {Li}, \citenamefont {Homes},\ and\
  \citenamefont {Zhu}}]{ma14}%
  \BibitemOpen
  \bibfield  {author} {\bibinfo {author} {\bibfnamefont {Chao}\ \bibnamefont
  {Ma}}, \bibinfo {author} {\bibfnamefont {Lijun}\ \bibnamefont {Wu}}, \bibinfo
  {author} {\bibfnamefont {Wei-Guo}\ \bibnamefont {Yin}}, \bibinfo {author}
  {\bibfnamefont {Huaixin}\ \bibnamefont {Yang}}, \bibinfo {author}
  {\bibfnamefont {Honglong}\ \bibnamefont {Shi}}, \bibinfo {author}
  {\bibfnamefont {Zhiwei}\ \bibnamefont {Wang}}, \bibinfo {author}
  {\bibfnamefont {Jianqi}\ \bibnamefont {Li}}, \bibinfo {author} {\bibfnamefont
  {C.~C.}\ \bibnamefont {Homes}}, \ and\ \bibinfo {author} {\bibfnamefont
  {Yimei}\ \bibnamefont {Zhu}},\ }\bibfield  {title} {\enquote {\bibinfo
  {title} {{Strong Coupling of the Iron-Quadrupole and Anion-Dipole
  Polarizations in {Ba}({Fe}$_{1-x}${Co}$_{x}$)$_{2}${As}$_{2}$}},}\ }\href
  {\doibase 10.1103/PhysRevLett.112.077001} {\bibfield  {journal} {\bibinfo
  {journal} {Phys. Rev. Lett.}\ }\textbf {\bibinfo {volume} {112}},\ \bibinfo
  {pages} {077001} (\bibinfo {year} {2014})}\BibitemShut {NoStop}%
\bibitem [{\citenamefont {Liu}\ \emph {et~al.}()\citenamefont {Liu},
  \citenamefont {Wu}, \citenamefont {Wu}, \citenamefont {Chen}, \citenamefont
  {Wang}, \citenamefont {Xie}, \citenamefont {Ying}, \citenamefont {Yan},
  \citenamefont {Li}, \citenamefont {Shi}, \citenamefont {Chu}, \citenamefont
  {Wu},\ and\ \citenamefont {Chen}}]{liu09}%
  \BibitemOpen
  \bibfield  {author} {\bibinfo {author} {\bibfnamefont {R.~H.}\ \bibnamefont
  {Liu}}, \bibinfo {author} {\bibfnamefont {T.}~\bibnamefont {Wu}}, \bibinfo
  {author} {\bibfnamefont {G.}~\bibnamefont {Wu}}, \bibinfo {author}
  {\bibfnamefont {H.}~\bibnamefont {Chen}}, \bibinfo {author} {\bibfnamefont
  {X.~F.}\ \bibnamefont {Wang}}, \bibinfo {author} {\bibfnamefont {Y.~L.}\
  \bibnamefont {Xie}}, \bibinfo {author} {\bibfnamefont {J.~J.}\ \bibnamefont
  {Ying}}, \bibinfo {author} {\bibfnamefont {Y.~J.}\ \bibnamefont {Yan}},
  \bibinfo {author} {\bibfnamefont {Q.~J.}\ \bibnamefont {Li}}, \bibinfo
  {author} {\bibfnamefont {B.~C.}\ \bibnamefont {Shi}}, \bibinfo {author}
  {\bibfnamefont {W.~S.}\ \bibnamefont {Chu}}, \bibinfo {author} {\bibfnamefont
  {Z.~Y.}\ \bibnamefont {Wu}}, \ and\ \bibinfo {author} {\bibfnamefont {X.~H.}\
  \bibnamefont {Chen}},\ }\bibfield  {title} {\enquote {\bibinfo {title} {A
  large iron isotope effect in {SmFeAsO}$_{1-x}${F}$_x$ and
  {Ba}$_{1-x}${K}$_x${Fe}$_2${As}$_2$},}\ }\href@noop {} {\ }\BibitemShut
  {NoStop}%
\bibitem [{\citenamefont {Singh}(1994)}]{singh}%
  \BibitemOpen
  \bibfield  {author} {\bibinfo {author} {\bibfnamefont {D.~J.}\ \bibnamefont
  {Singh}},\ }\href@noop {} {\emph {\bibinfo {title} {Planewaves,
  Pseudopotentials and the LAPW method}}}\ (\bibinfo  {publisher} {Kluwer
  Adademic},\ \bibinfo {address} {Boston},\ \bibinfo {year} {1994})\BibitemShut
  {NoStop}%
\bibitem [{\citenamefont {Singh}(1991)}]{singh91}%
  \BibitemOpen
  \bibfield  {author} {\bibinfo {author} {\bibfnamefont {David}\ \bibnamefont
  {Singh}},\ }\bibfield  {title} {\enquote {\bibinfo {title} {{Ground-state
  properties of lanthanum: Treatment of extended-core states}},}\ }\href
  {\doibase 10.1103/PhysRevB.43.6388} {\bibfield  {journal} {\bibinfo
  {journal} {Phys. Rev. B}\ }\textbf {\bibinfo {volume} {43}},\ \bibinfo
  {pages} {6388--6392} (\bibinfo {year} {1991})}\BibitemShut {NoStop}%
\bibitem [{wie()}]{wien2k}%
  \BibitemOpen
  \href@noop {} {}\bibinfo {note} {P. Blaha, K. Schwarz, G.~K.~H. Madsen, D.
  Kvasnicka and J. Luitz, WIEN2k, {\it An augmented plane wave plus local
  orbitals program for calculating crystal properties} (Techn.
  Universit{\"{a}}t Wien, Austria, 2001).}\BibitemShut {Stop}%
\bibitem [{pho()}]{phonon}%
  \BibitemOpen
  \href@noop {} {}\bibinfo {note} {K. Parlinksi, Software PHONON
  (2003).}\BibitemShut {Stop}%
\bibitem [{\citenamefont {Gretarsson}\ \emph {et~al.}(2011)\citenamefont
  {Gretarsson}, \citenamefont {Lupascu}, \citenamefont {Kim}, \citenamefont
  {Casa}, \citenamefont {Gog}, \citenamefont {Wu}, \citenamefont {Julian},
  \citenamefont {Xu}, \citenamefont {Wen}, \citenamefont {Gu}, \citenamefont
  {Yuan}, \citenamefont {Chen}, \citenamefont {Wang}, \citenamefont {Khim},
  \citenamefont {Kim}, \citenamefont {Ishikado}, \citenamefont {Jarrige},
  \citenamefont {Shamoto}, \citenamefont {Chu}, \citenamefont {Fisher},\ and\
  \citenamefont {Kim}}]{gretarsson11}%
  \BibitemOpen
  \bibfield  {author} {\bibinfo {author} {\bibfnamefont {H.}~\bibnamefont
  {Gretarsson}}, \bibinfo {author} {\bibfnamefont {A.}~\bibnamefont {Lupascu}},
  \bibinfo {author} {\bibfnamefont {Jungho}\ \bibnamefont {Kim}}, \bibinfo
  {author} {\bibfnamefont {D.}~\bibnamefont {Casa}}, \bibinfo {author}
  {\bibfnamefont {T.}~\bibnamefont {Gog}}, \bibinfo {author} {\bibfnamefont
  {W.}~\bibnamefont {Wu}}, \bibinfo {author} {\bibfnamefont {S.~R.}\
  \bibnamefont {Julian}}, \bibinfo {author} {\bibfnamefont {Z.~J.}\
  \bibnamefont {Xu}}, \bibinfo {author} {\bibfnamefont {J.~S.}\ \bibnamefont
  {Wen}}, \bibinfo {author} {\bibfnamefont {G.~D.}\ \bibnamefont {Gu}},
  \bibinfo {author} {\bibfnamefont {R.~H.}\ \bibnamefont {Yuan}}, \bibinfo
  {author} {\bibfnamefont {Z.~G.}\ \bibnamefont {Chen}}, \bibinfo {author}
  {\bibfnamefont {N.-L.}\ \bibnamefont {Wang}}, \bibinfo {author}
  {\bibfnamefont {S.}~\bibnamefont {Khim}}, \bibinfo {author} {\bibfnamefont
  {K.~H.}\ \bibnamefont {Kim}}, \bibinfo {author} {\bibfnamefont
  {M.}~\bibnamefont {Ishikado}}, \bibinfo {author} {\bibfnamefont
  {I.}~\bibnamefont {Jarrige}}, \bibinfo {author} {\bibfnamefont
  {S.}~\bibnamefont {Shamoto}}, \bibinfo {author} {\bibfnamefont {J.-H.}\
  \bibnamefont {Chu}}, \bibinfo {author} {\bibfnamefont {I.~R.}\ \bibnamefont
  {Fisher}}, \ and\ \bibinfo {author} {\bibfnamefont {Young-June}\ \bibnamefont
  {Kim}},\ }\bibfield  {title} {\enquote {\bibinfo {title} {Revealing the dual
  nature of magnetism in iron pnictides and iron chalcogenides using x-ray
  emission spectroscopy},}\ }\href {\doibase 10.1103/PhysRevB.84.100509}
  {\bibfield  {journal} {\bibinfo  {journal} {Phys. Rev. B}\ }\textbf {\bibinfo
  {volume} {84}},\ \bibinfo {pages} {100509(R)} (\bibinfo {year}
  {2011})}\BibitemShut {NoStop}%
\end{thebibliography}

%\newpage
%\hspace*{-4.0cm}
\begin{figure}[t]
%\centerline{\includegraphics[width=7.0in]{SuppMat.ps}}%
\vspace*{-1.0cm}
\hspace*{-1.80cm}
\includegraphics[width=8.40in]{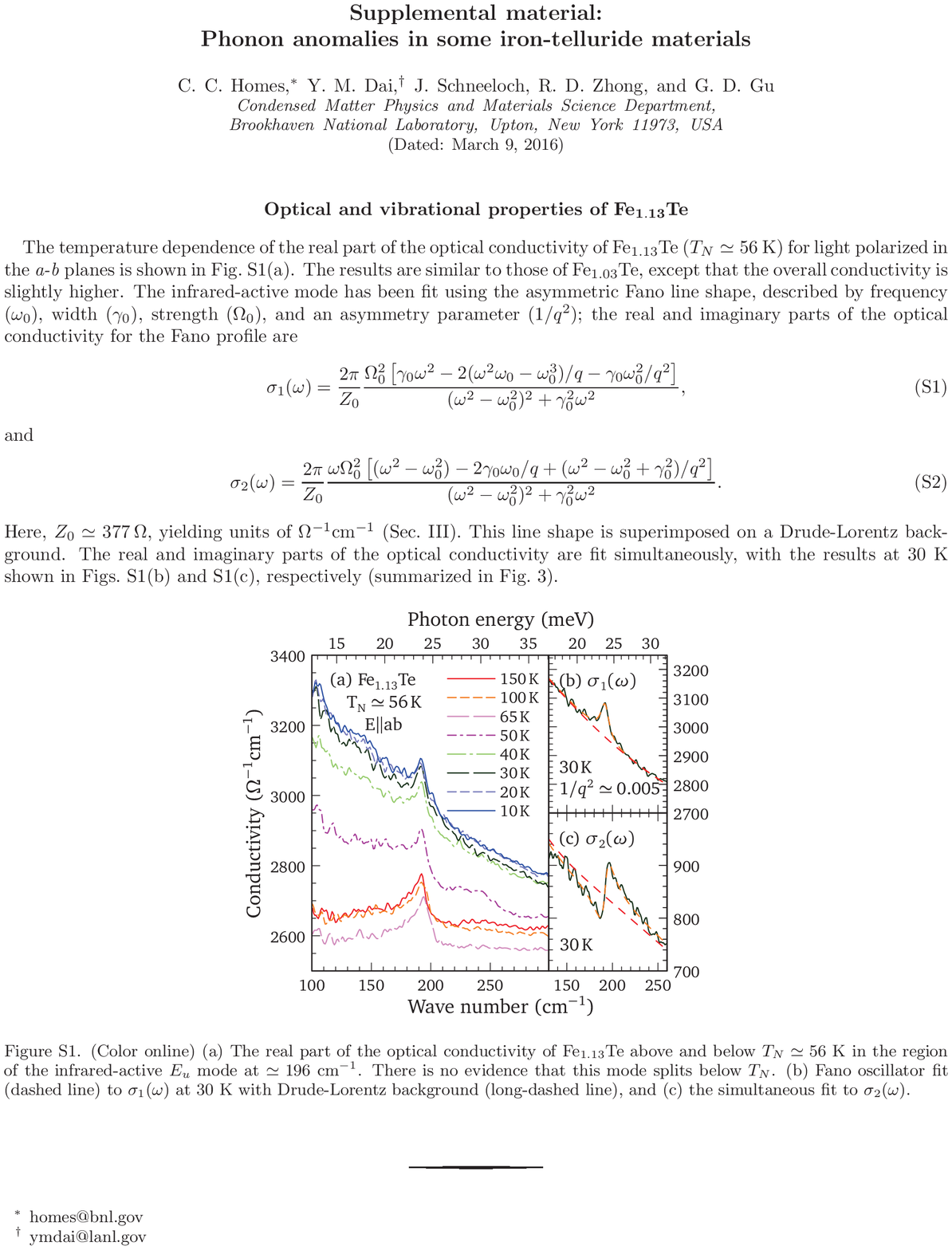}
\end{figure}

\end{document}